\documentclass[11pt,draftcls,onecolumn]{IEEEtran}
\usepackage{amsmath,graphicx}
\usepackage{amsmath,graphicx}
\usepackage{enumerate}
\usepackage{amsbsy}
\usepackage{amssymb}
\usepackage{amsthm}
\usepackage{amscd}
\usepackage{subfigure}
\usepackage{color}
\def\bSig{{\boldsymbol{\Sigma}}}

\def\balpha{{\boldsymbol{\alpha}}}
\def\bzero{{\boldsymbol{0}}}
\def\bone{{\boldsymbol{1}}}
\def\c1{{\textcircled{a}}}
\def\ba{{\boldsymbol{a}}}
\def\bb{{\boldsymbol{b}}}
\def\bc{{\boldsymbol{c}}}

\def\be{{\boldsymbol{e}}}

\def\bn{{\boldsymbol{n}}}

\def\bp{{\boldsymbol{p}}}

\def\bs{{\boldsymbol{s}}}

\def\bu{{\boldsymbol{u}}}
\def\bv{{\boldsymbol{v}}}
\def\bw{{\boldsymbol{w}}}
\def\bx{{\boldsymbol{x}}}
\def\by{{\boldsymbol{y}}}

\def\bB{{\boldsymbol{B}}}

\def\bD{{\boldsymbol{D}}}
\def\bE{{\boldsymbol{E}}}

\def\bI{{\boldsymbol{I}}}
\def\bJ{{\boldsymbol{J}}}

\def\bM{{\boldsymbol{M}}}

\def\bP{{\boldsymbol{P}}}
\def\bQ{{\boldsymbol{Q}}}
\def\bR{{\boldsymbol{R}}}
\def\bS{{\boldsymbol{S}}}

\def\bU{{\boldsymbol{U}}}

\def\bX{{\boldsymbol{X}}}

\def\complexC{{\mathbb{C}}}
\def\realR{{\mathbb{R}}}

\def\bzero{{\boldsymbol{0}}}
\def\bone{{\boldsymbol{1}}}

\newtheorem*{Theorem1}{Theorem 1}
\newtheorem*{Theorem2}{Theorem 2}
\newtheorem*{Theorem3}{Theorem 3}
\newtheorem*{Theorem4}{Theorem 4}
\newtheorem*{Theorem5}{Theorem 5}
\newtheorem*{Theorem6}{Theorem 6}
\newtheorem*{Theorem7}{Theorem 7}

\title{Designing Unimodular Codes via Quadratic Optimization is not Always Hard}
\author{
Mojtaba~Soltanalian* and Petre~Stoica,~\IEEEmembership{Fellow,~IEEE}
\thanks{This work was supported in part by the European Research Council (ERC) under Grant \#228044 and the Swedish Research Council. The authors are with the Dept. of Information Technology, Uppsala University, Uppsala, SE 75105, Sweden.

 * Please address all the correspondence to Mojtaba Soltanalian, Phone: (+46) 18-471-3168; Fax: (+46) 18-511925; Email: mojtaba.soltanalian@it.uu.se}% <-this % stops a space
}
\begin{document}
%\ninept
%
\maketitle
\begin{abstract}
The NP-hard problem of optimizing a quadratic form over the unimodular vector set arises in radar code design scenarios as well as other active sensing and communication applications. To tackle this problem (which we call unimodular quadratic programming (UQP)), several computational approaches are devised and studied. A specialized local optimization scheme for UQP is introduced and shown to yield superior results compared to general local optimization methods. Furthermore, a  \textbf{m}onotonically \textbf{er}ror-bound \textbf{i}mproving \textbf{t}echnique (MERIT) is proposed to obtain the global optimum or a local optimum of UQP with good sub-optimality guarantees. The provided sub-optimality guarantees are case-dependent and generally outperform the $\pi/4$ approximation guarantee of semi-definite relaxation. Several numerical examples are presented to illustrate the performance of the proposed method. The examples show that for  cases including several matrix structures used in radar code design, MERIT can solve UQP efficiently in the sense of sub-optimality guarantee and computational time.
%[The word "Hard" is used such that it does not fall directly in the rigorous context of complexity. We include some class of matrices which can be handled easily by the proposed algorithm.]
\end{abstract}
\begin{keywords}
radar codes, unimodular codes, quadratic programming. %Constant modulus, NP-hard,
\end{keywords}
\section{Introduction} \label{sec:intro}
Unimodular codes are used in many active sensing and communication systems mainly as a result of the their optimal (i.e. unity) peak-to-average-power ratio (PAR). The design of such codes can be often formulated as the optimization of a quadratic form (see sub-section \ref{subsec:background} for examples). Therefore, we will study the problem
\begin{eqnarray} \label{eq:UQP}
\mbox{UQP: }~ \max_{\bs \in \Omega^n} \bs^H \bR \bs
\end{eqnarray}
where $\bR \in \complexC^{n \times n}$ is a given Hermitian matrix, $(.)^H$ denotes the vector/matrix Hermitian transpose, $\Omega$ represents the unit circle, i.e.  
%\begin{eqnarray}
$\Omega=\{s \in \complexC : ~| s | =1 \}$ 
%\end{eqnarray}
and UQP stands for Unimodular Quadratic Program(ming).

%The problem is shown to be NP-hard []. 

\subsection{Motivating Applications} \label{subsec:background}
To motivate the UQP formulation considered above, we present four scenarios in which a design problem in active sensing or communication boils down to an UQP. %It is worth noting that the applications of UQP could be rather wider than the examples we consider in this paper. The three scenarios are as follows:

$\bullet$~  \emph{Designing codes that optimize the SNR or the CRLB:}
We consider a monostatic radar which transmits a linearly encoded burst of pulses. The observed backscattered signal $\bv$ can be written as (see, e.g. \cite{DeMaio-similarity}):
\begin{eqnarray}
\bv= a (\bc \odot \bp) + \bw,
\end{eqnarray}
where $a$ represents both channel propagation and backscattering effects, 
$\bw$ is the disturbance/noise component, $\bc$ is the unimodular vector containing the code elements, $\bp=( 1, e^{j2\pi f_d T_r}, \cdots, e^{j2\pi (n-1) f_d T_r})^T$ is the temporal steering vector with $f_d$ and $T_r$ being the target Doppler frequency and pulse repetition time, respectively, and the symbol $\odot$ stands for the Hadamard (element-wise) product of matrices. 

Under the assumption that $\bw$ is a zero-mean complex-valued circular Gaussian vector with known positive definite covariance matrix $\mathbb{E}[\bw \bw^H]=\bM$, the signal-to-noise ratio (SNR) is given by \cite{DeMaio-selection}
\begin{eqnarray} \label{eq:SNR}
\mbox{SNR}=| a |^2 \bc^H \bR \bc
\end{eqnarray}  
where $\bR = \bM^{-1} \odot (\bp \bp^H)^*$ with  $(.)^*$ denoting the vector/matrix complex conjugate. Therefore, the problem of designing codes optimizing the SNR of the radar system can be formulated directly as an UQP.  Additionally, the Cramer-Rao lower bound (CRLB) for the target Doppler frequency estimation (which yields a lower bound on the variance of any unbiased target Doppler frequency estimator) is given by \cite{DeMaio-selection}
\begin{eqnarray}
\mbox{CRLB}&=&\left( 2 | a |^2 (\bc \odot \bp \odot \bu)^H \bM^{-1} (\bc \odot \bp \odot \bu) \right)^{-1} \\ \nonumber
&=& \left( 2 | a |^2 \bc^H \bR' \bc \right)^{-1}
\end{eqnarray}
where $\bu=(0, j 2 \pi T_r, \cdots, j 2 \pi (n-1) T_r )^T$ and 
%\begin{eqnarray}
$\bR' = \bM^{-1} \odot (\bp \bp^H)^* \odot (\bu \bu^H)^*$.
%\end{eqnarray} 
Therefore the minimization of CRLB can also be formulated as an UQP. For the simultaneous optimization of SNR and CRLB see \cite{DeMaio-selection}.
\vspace{5pt}

$\bullet$~  \emph{Synthesizing cross ambiguity functions (CAFs):}
The ambiguity function (which is widely used in active sensing applications \cite{seq_book}\cite{levanon}) represents the two-dimensional response of the matched filter to a signal with time delay $\tau$ and Doppler frequency shift $f$. The more general concept of cross ambiguity function occurs when the match filter is replaced by a mismatched filter. The cross ambiguity function (CAF) is defined as
\begin{eqnarray} \label{eq:CAF}
\chi(\tau,f)=\int_{-\infty}^{\infty} u(t) v^*(t+\tau) e^{j 2 \pi f t} dt
\end{eqnarray}
where $u(t)$ and $v(t)$ are the transmit signal and the receiver filter, respectively (the ambiguity function is obtained from (\ref{eq:CAF}) with $v(t) = u(t)$). In several applications $u(t)$ and $v(t)$ are given by:
\begin{eqnarray}
\label{eqWaveDef}
u(t) = \sum_{k=1}^n x_k p_k(t),~~~v(t) = \sum_{k=1}^n y_k p_k(t)
\end{eqnarray}
where $\{p_k(t)\}$ are pulse-shaping functions (with the rectangular pulse as a common example), and 
\begin{eqnarray}
\bx=( x_1 \cdots x_n )^T,~~\by=( y_1 \cdots y_n)^T
\end{eqnarray}
are the code and, respectively, the filter vectors. The design problem of synthesizing a desired CAF has a small number of free variables (i.e. the entries of the vectors $\bx$ and $\by$)  compared to the large number of constraints arising from two-dimensional matching criteria (to a given $|\chi(\tau,f)|$). Therefore, the problem is generally considered to be difficult and there are not many methods to synthesize a desired (cross) ambiguity function. Below, we describe briefly the cyclic approach of \cite{Hao_CAF} for CAF design.

%Define the matrix $\bJ(\tau,f)$ as 
%\begin{eqnarray}
%[\bJ(\tau,f)]_{k,l}= \int_{-\infty}^{\infty} p_l(t) p_k^*(t+\tau) e^{j2\pi f t} \ dt, ~~~ 1 \leq k,l \leq n.
%\end{eqnarray}
%Then $\chi(\tau,f) = \by^H \bJ(\tau,f) \bx $, and as a result 
The problem of matching a desired $|\chi(\tau,f)| = d(\tau,f)$ can be formulated as the minimization of the criterion \cite{Hao_CAF}
\begin{eqnarray} \label{eq:CAF_crit}
g(\bx,\by,\phi)= \int_{-\infty}^{\infty} \int_{-\infty}^{\infty} w(\tau,f) \left|  d(\tau,f) e^{j \phi(\tau,f)} - \by^H \bJ(\tau,f) \bx  \right|^2 \ d\tau df 
\end{eqnarray}
where $\bJ(\tau,f) \in \complexC^{n \times n}$ is given, $w(\tau,f)$ is a weighting function that specifies the CAF area which needs to be emphasized and $\phi(\tau,f)$ represent auxiliary phase variables. It is not difficult to see that for fixed $\bx$ and $\by$, the minimizer $\phi(\tau,f)$ is given by 
%\begin{eqnarray}
$\phi(\tau,f)= \arg \{ \by^H \bJ(\tau,f) \bx  \}$. 
%\end{eqnarray}
For fixed $\phi(\tau,f)$ and $\bx$, the criterion $g$ can be written as
\begin{eqnarray}
g(\by)&=&\by^H \bD_1 \by - \by^H \bB^H \bx - \bx^H \bB \by + \mbox{const}_1 \\ \nonumber
&=& (\by - \bD_1^{-1} \bB^H \bx)^H \bD_1  (\by - \bD_1^{-1} \bB^H \bx) + \mbox{const}_2
\end{eqnarray}
%where
%\begin{eqnarray}
%\bD_1= \int_{-\infty}^{\infty} \int_{-\infty}^{\infty} w(\tau,f)  \bJ^H(\tau,f) \bx \bx^H \bJ(\tau,f)  \ d\tau df ,\\ \nonumber
%\bB= \int_{-\infty}^{\infty} \int_{-\infty}^{\infty} w(\tau,f)  d(\tau,f) e^{j \phi(\tau,f)} \bJ^H(\tau,f) \ d\tau df.
%\end{eqnarray} 
where $\bB$ and $\bD_1$ are given matrices in $\complexC^{n \times n}$ \cite{Hao_CAF}. Due to practical considerations, the transmit coefficients $\{x_k\}$  must have low PAR values. However, the receiver coefficients $\{y_k\}$ need not be constrained in such a way. Therefore, the minimizer $\by$ of $g(\by)$ is given by $\by=\bD_1^{-1} \bB^H \bx$. Similarly, for fixed $\phi(\tau,f)$ and $\by$, the criterion $g$ can be written as
\begin{eqnarray} \label{eq:CAF_UQP1}
g(\bx)= \bx^H \bD_2 \bx - \bx^H \bB \by - \by^H \bB^H \bx + \mbox{const}_3
\end{eqnarray}
%where
%\begin{eqnarray}
%\bD_2= \int_{-\infty}^{\infty} \int_{-\infty}^{\infty} w(\tau,f)  \bJ^H(\tau,f) \by \by^H \bJ(\tau,f)  \ d\tau df.
%\end{eqnarray}
where $\bD_2 \in \complexC^{n \times n}$ is given \cite{Hao_CAF}. If a unimodular code vector $\bx$ is desired then the optimization of $g(\bx)$ is an UQP as $g(\bx)$ can be written as
\begin{eqnarray} \label{eq:CAF_UQP2}
g(\bx)= \left( \begin{array}{c}
e^{j \varphi} \bx  \\
e^{j \varphi} \\ 
\end{array} \right)^H  
\left( \begin{array}{cc}
\bD_2 & -\bB \by \\
-(\bB \by)^H & 0 \\ 
\end{array} \right)\left( \begin{array}{c}
e^{j \varphi} \bx  \\
e^{j \varphi} \\ 
\end{array} \right)+ \mbox{const}_3
\end{eqnarray}
where $\varphi \in [0, 2 \pi)$ is a free phase variable.
\vspace{5pt}

$\bullet$~ \emph{Steering vector estimation in adaptive beamforming:} Consider a linear array with $n$ antennas. The output of the array at time instant $k$ can be expressed as \cite{Petre-beamformig}
\begin{eqnarray} 
\bx_k= s_k \ba +\bn_k
\end{eqnarray}
with $\{s_k\}$ being the signal waveform, $\ba$ the associated steering vector (with $|[\ba]_l|=1$, $1 \leq l \leq n$), and $\bn_k$ the vector accounting for all independent interferences.

%For a given weight (beamforming) vector $\bw$, the beamformer output (at the time instant $k$) is given by
%$%\begin{eqnarray}
%y_k=\bw^H \bx_k$. 
%%\end{eqnarray}
%Assuming a precise knowledge of the steering vector, the optimal weight vector is obtained by maximizing the output SNR,
%\begin{eqnarray} \label{eq:beam-snr}
%\mbox{SNR}= \frac{\sigma_s^2 | \bw^H \ba |^2}{\bw^H \bR_\bn \bw},
%\end{eqnarray}  
%where $\sigma_s^2$ is the signal power and $\bR_\bn=\mathbb{E}[\bn_k \bn^H_k]$. Note that in practice $\bR_\bn$ is estimated using the sample covariance matrix 
%$%\begin{eqnarray}
%\widehat{\bR}= \frac{1}{T} \sum_{k=1}^{T} \bx_k \bx^H_k
%$ %\end{eqnarray}
%where $T$ is the number of training data samples. 

%%For cases in which the entries of $\bw$ are not constrained (e.g. to the unimodular alphabet), 
%The maximization of the SNR function in (\ref{eq:beam-snr}) with respect to  $\bw \in \complexC^{n}$ yields the following well-known closed-from expression \cite{Petre-beamformig}\nocite{Khabbazibasmenj}:
%\begin{eqnarray}
%\bw=\bR_\bn^{-1} \ba.
%\end{eqnarray} 
%However 
The true steering vector is usually unknown in practice, and it can therefore be considered as an unimodular vector to be determined \cite{steer-unique}. Define the sample covariance matrix of $\{\bx_k\}$ as
$%\begin{eqnarray}
\widehat{\bR}= \frac{1}{T} \sum_{k=1}^{T} \bx_k \bx^H_k
$ %\end{eqnarray}
where $T$ is the number of training data samples. Assuming some prior knowledge on $\ba$ (which can be represented by $\arg(\ba)$ being in a given sector $\Theta$), the problem of estimating the steering vector can be formulated as \cite{Khabbazibasmenj}
\begin{eqnarray} \label{eq:UQP-related}
\min_{\ba}~~ \ba^H \widehat{\bR}^{-1} \ba~ \\ \nonumber
\mbox{s.t. } \arg(\ba) \in \Theta,
\end{eqnarray}
hence an UQP-type problem. Such problems can be tackled using general local optimization techniques or the optimization scheme introduced in Section \ref{sec:local}.
%\vspace{5pt}

$\bullet$~  \emph{Maximum likelihood (ML) detection of unimodular codes:}
Assume the linear model
\begin{eqnarray}
\by = \bQ \bs + \bn
\end{eqnarray}
where $\bQ$ represents a multiple-input multiple-output (MIMO) channel, $\by$ is the received signal, $\bn$ is the additive white Gaussian noise and $\bs$ contains the unimodular symbols which are to be estimated. The ML detection of $\bs$ may be stated as
\begin{eqnarray}
\widehat{\bs}_{ML}= \arg \min_{\bs \in \Omega^n} \| \by - \bQ \bs \|_2
\end{eqnarray}
It is straightforward to verify that the above optimization problem is equivalent to the UQP \cite{Jalden}:
\begin{eqnarray}
\min_{\overline{\bs} \in \Omega^{n+1}}~ \overline{\bs}^H \bR \overline{\bs}
\end{eqnarray}
where
\begin{eqnarray}
\bR=\left( \begin{array}{cc}
\bQ^H \bQ & -\bQ^H \by \\
-\by^H \bQ & 0 \\
\end{array} \right) \mbox{, } \overline{\bs} = \left( \begin{array}{c}
e^{j \varphi} \bs \\
e^{j \varphi}  \\
\end{array} \right)
\end{eqnarray}
and $\varphi \in [0, 2 \pi)$ is a free phase variable.

%
%Theorem 1 compares the hardness of UQP with any related $M$-UQP in terms of the number of local optimas.

%\begin{Theorem1} \label{th:hardness-compare}
%The UQP problem in () has at least as many local optimas as the $M$-UQP in () for all $M \geq 2$.
%\end{Theorem1}
%\begin{proof}
%Note that $\bs \in  \Omega_M$ is a local optima of $M$-UQP in (), if and only if, for all vector pairs
%\begin{eqnarray}
%\left\{ \begin{array}{l}
%\bs_{k,-}=e^{j \left( \arg(\bs) - \frac{2 \pi}{M} \be_k \right)}\\
%\bs_{k,+}=e^{j \left( \arg(\bs) + \frac{2 \pi}{M} \be_k \right)} 
% \end{array} \right. , ~ k \in \{ 1,2,\cdots,n\}
%\end{eqnarray}
%we have that
%\begin{eqnarray} \label{eq:MUQP-local}
%\left\{ \begin{array}{l}
%\bs^H \bR \bs \geq \bs_{k,-}^H \bR \bs_{k,-}, \\
%\bs^H \bR \bs \geq \bs_{k,+}^H \bR \bs_{k,+}.
%\end{array} \right.
%\end{eqnarray}
%Therefore, as $\Omega_M \subset \Omega$, the latter result implies that there exist at least one $\bs= \bs' \in  \Omega$ satisfying (\ref{eq:MUQP-local}) for which
%\begin{eqnarray} \label{eq:phase-limit}
%\arg(\bs_{k,-}) \prec \arg(\bs') \prec \arg(\bs_{k,+}), \\ \nonumber
%\forall ~k \in \{ 1,2,\cdots,n\}.
%\end{eqnarray}
%Since $\bs'^H \bR \bs'$ is upper bounded (by $n \sigma_1$ where $\sigma_1$ is the maximum eigenvalue of $R$), the set 
%\begin{eqnarray}
%\{ \bs'^H \bR \bs':~ \bs'\in  \Omega \mbox{ satisfies (\ref{eq:phase-limit}).} \}
%\end{eqnarray}
%has a supremum. Not OK!
%\end{proof} 
%\vspace{5pt}

\subsection{Related Work} \label{subsec:previous}
In \cite{zhang-complexquad}, the NP-hardness of UQP is proven by employing a reduction from an NP-complete matrix partitioning problem. The UQP in (\ref{eq:UQP}) is often studied along with the following (still NP-hard) related problem in which the decision variables are discrete:
\begin{eqnarray} \label{eq:MUQP}
\mbox{$m$-UQP:}~ \max_{\bs \in \Omega^n_m} \bs^H \bR \bs 
\end{eqnarray}
where $\Omega_m=\{ 1, e^{j\frac{2 \pi}{m}},\cdots,e^{j\frac{2 \pi}{m}(m-1)}  \}$. Note that the latter problem coincides with the UQP in (\ref{eq:UQP}) as $m \rightarrow \infty$. The authors of \cite{low-rank} show that when the matrix $\bR$ is rank-deficient (more precisely, when $d=$rank$(\bR)$ behaves like $\mathcal{O}(1)$ with respect to the problem dimension) the $m$-UQP problem can be solved in polynomial-time and they propose a $\mathcal{O}((mn/2)^{2d})$-complexity algorithm to solve (\ref{eq:MUQP}). However, such algorithms are not applicable to the UQP which corresponds to an infinite $m$.  

Studies on polynomial-time algorithms for UQP (and $m$-UQP) have been extensive (e.g. see [9]-[19] and the references therein). In particular, the semi-definite relaxation (SDR) technique has been one of the most appealing approaches to the researchers. To derive an SDR, we note that \nocite{sergio-multiusercomplexity}\nocite{MaVo-blindML}\nocite{Cui_OFDM_spheredecoing}
%\begin{eqnarray}
$\bs^H \bR \bs =  \mbox{tr} (\bs^H \bR \bs) =  \mbox{tr} (\bR \bs \bs^H )$. 
%\end{eqnarray}
Hence, the UQP can be rewritten as
\begin{eqnarray} \label{eq:rUQP0}
\max_{\bS} \,  \mbox{tr} (\bR \bS ) ~~~~~~~ \\ \nonumber
\mbox{s.t.  } \bS= \bs \bs^H,~ \bs \in \Omega^n.
\end{eqnarray}
If we relax (\ref{eq:rUQP0}) by removing the rank constraint on $\bS$ and the unimodularity constraint on $\bs$ then the result is a semi-definite program:
\begin{eqnarray} \label{eq:rUQP}
\mbox{SDP: }~ \max_{\bS}  \mbox{tr} (\bR \bS )~~~~~~~ \\ \nonumber
\mbox{s.t. } [\bS]_{k,k}=1,~~ 1 \leq k \leq n, ~~~\\ \nonumber
~~~~~~ \bS \mbox{ is positive semi-definite.}
\end{eqnarray}
The above SDP can be solved in polynomial time using interior-point methods \cite{convex_boyd}. The approximation of the UQP solution based on the SDP solution can be accomplished in several ways. For example, we can approximate the phase values of the solution $\bs$ using a rank-one approximation of $\bS$. A more effective approach for guessing $\bs$ is based on randomized approximations (see \cite{zhang-complexquad}, \cite{Goemans}  and \cite{DeMaio-PAR}). A detailed guideline for randomized approximation of the UQP solution can be found in \cite{DeMaio-PAR}. In addition, we refer the interested reader to the survey of the rich literature on SDR in \cite{SDR_mag}. 

Analytical assessments of the quality of the UQP solutions obtained by SDR and randomized approximation are available. Let $v_{SDR}$ be the expected value of the UQP objective at the obtained randomized solution. Let $v_{opt}$ represent the optimal value of the UQP objective. We have
\begin{eqnarray}
\gamma v_{opt} \leq v_{SDR} \leq v_{opt}
\end{eqnarray}
with the sub-optimality guarantee coefficient $\gamma=\pi /4$  \cite{zhang-complexquad}\cite{so-approxSDR}. Note that the sub-optimality coefficient of the solution obtained by SDR can be arbitrarily close to $\pi /4$ (e.g., see \cite{so-approxSDR}).

% In the following, we describe the technique as the main tool  UQP (and $M$-UQP) 

%Based on the complexity of such problems, there has been an extensive study on the

\subsection{Contributions of this Work} \label{subsec:contributions}
Besides SDR, the literature does not offer many other numerical approaches to tackle UQP. In this paper, a specialized local optimization scheme for UQP is proposed. The proposed computationally efficient local optimization approach can be used to tackle UQP as well as improve upon the solutions obtained by other methods such as SDR. Furthermore, a \textbf{m}onotonically \textbf{er}ror-bound \textbf{i}mproving \textbf{t}echnique (called MERIT) is introduced to obtain the global optimum or a local optimum of UQP with good sub-optimality guarantees. Note that:
\begin{itemize}
\item MERIT provides case-dependent sub-optimality guarantees. To the best of our knowledge, such guarantees for UQP were not known prior to this work. Using the proposed method one can generally obtain better performance guarantees compared to the analytical worst-case guarantees (such as $\gamma=\pi /4$ for SDR).
\item The provided case-dependent sub-optimality guarantees are of practical importance in decision making scenarios. For instance in some cases the UQP solution obtained by SDR (or other optimization methods) might achieve good objective values. %with respect to the sub-optimality metric. 
However, unless the goodness of the obtained solution is known (this goodness can be determined using the proposed bounds), the solution cannot be trusted.
%\item The optimal objective of the UQP or a tight sub-optimality guarantee can be determined via MERIT. %The provided approximations of the optimal UQP objective can be used to examine the sub-optimality of the solutions obtained by any other UQP optimization scheme.
\item Using MERIT, numerical evidence is provided to show that several UQPs (particularly those which occur in active sensing code design) can be solved efficiently without sacrificing the solution accuracy.
\end{itemize}

Finally, we believe that the general ideas of this work can be adopted to tackle $m$-UQP as the finite alphabet case of UQP. However, a detailed study of $m$-UQP is beyond the scope of this paper.

%A method verifying the global optimality of any given solution $\bs$ to UQP is missing in the literature. In this paper, we provide a polynomial-time algorithm that fulfills such verification. As a result, we show that UQP is NP.  In combination with the previous results showing the NP-hardness of UQP, our results imply that UQP is indeed an NP-complete problem and therefore improves our knowledge regarding the complexity of UQP.

\subsection{Organization of the Paper} \label{subsec:organization}

The rest of this work is organized as follows. Section \ref{sec:properties} discusses several properties of UQP. Section \ref{sec:local} introduces a specialized local optimization method. Section \ref{sec:cone} presents a cone approximation that is used in Section \ref{sec:global} to derive the algorithmic form of MERIT for UQP. Several numerical examples are provided in section \ref{sec:numerical}. Finally, Section \ref{sec:conclusion} concludes the paper.

\emph{Notation:} We use bold lowercase letters for vectors/sequences and bold uppercase letters for matrices. $(.)^T$ denotes the vector/matrix transpose. $\bone$ and $\bzero$ are the all-one and all-zero vectors/matrices. $\be_k$ is the $k^{th}$ standard basis vector in $\complexC^n$.  $\| \bx \|_n$ or the $l_n$-norm of the vector $\bx$ is defined as $\left( \sum_k |\bx(k)|^n \right)^\frac{1}{n}$ where  $\{ \bx(k) \}$ are the entries of $\bx$. The Frobenius norm of a matrix $\bX$ (denoted by  $\| \bX \|_F$) with entries $\{ \bX(k,l) \}$ is equal to $\left( \sum_{k,l} |\bX(k,l)|^2 \right)^\frac{1}{2}$.  We use $\Re(\bX)$ to denote the matrix obtained by collecting the real parts of the entries of $\bX$. The matrix $e^{j \bX}$ is defined element-wisely as $\left[e^{j \bX}\right]_{k,l}=e^{j [\bX]_{k,l}}$.  $\arg(.)$ denotes the phase angle (in radians) of the vector/matrix argument. $\mathbb{E}[.]$ stands for the expectation operator. $\mathbf{Diag}(.)$  denotes the diagonal matrix formed by the entries of the vector argument, whereas  $\mathbf{diag}(.)$ denotes the vector formed by collecting the diagonal entries of the matrix argument. $\sigma_k(\bX)$ represents the $k^{th}$ maximal eigenvalue of $\bX$. Finally,  $\realR$ and $ \complexC $ represent the set of real and complex numbers, respectively. %for any real number $x$, the function $[x]$ yields the closest integer to $x$ (the largest is chosen when this integer is not unique). 

\section{Some Properties of UQP}\label{sec:properties}
In this section, we study several properties of UQP. The discussed properties lay the grounds for a better understanding of UQP as well as the tools proposed to tackle it in the following sections.

\subsection{Basic Properties} \label{subsec:basic}
The UQP formulation in (\ref{eq:UQP}) covers both maximization and minimization of quadratic forms (one can obtain the minimization of the quadratic form in (\ref{eq:UQP}) by considering $-\bR$ in lieu of $\bR$). In addition, without loss of generality, the Hermitian matrix $\bR$ can be assumed to be positive (semi)definite. If $\bR$ is not positive (semi)definite, we can make it so using the diagonal loading technique (i.e. 
%\begin{eqnarray}
$\bR \leftarrow \bR + \lambda \bI$ 
%\end{eqnarray}
where $\lambda \geq -\sigma_n(\bR )$). Note that such a diagonal loading does not change the solution of UQP as $\bs^H (\bR + \lambda \bI) \bs = \bs^H  \bR \bs + \lambda n$. Next, we note that if $\widetilde{\bs}$ is a solution to UQP then $e^{j \phi} \widetilde{\bs}$ (for any $\phi \in [0, 2 \pi)$) is also a valid solution. %In particular, for real-valued $\bR$, $\widetilde{\bs}^*$ also yields a valid solution.
 To establish connections among different UQPs, Theorem 1 presents a bijection among the set of matrices leading to the same solution.

\begin{Theorem1} \label{th:K(s)}
Let $\mathcal{K}(\bs) $ represent the set of matrices $\bR$ for which a given $\bs \in \Omega^n$ is the global optimizer of UQP. Then
\begin{enumerate}
\item $\mathcal{K}(\bs)$ is a convex cone.
\item For any two vectors $\bs_1,\bs_2  \in \Omega^n$, the one-to-one mapping (where $\bs_0=\bs_1^* \odot \bs_2$)
\begin{eqnarray}
\bR \in \mathcal{K}(\bs_1) \Longleftrightarrow  \bR \odot (\bs_0 \bs_0^H) \in   \mathcal{K}(\bs_2)
\end{eqnarray}
holds among the matrices in $\mathcal{K}(\bs_1)$ and $\mathcal{K}(\bs_2)$.
%There exists a one-to-one mapping among $\mathcal{K}(\bs)$ for all $\bs \in \Omega$, via the relationship
\end{enumerate}
\end{Theorem1}
\begin{proof}
See the Appendix.
\end{proof}
\vspace{5pt}
It is interesting to note that in light of the above result, the characterization of the cone $\mathcal{K}(\bs)$ for any given $\bs=\widetilde{\bs}$ leads to a complete characterization of all $\mathcal{K}(\bs)$, $\bs \in \Omega^n$, and thus solving any UQP. However, the NP-hardness of UQP suggests that such a characterization cannot be expected. Further discussions regarding the characterization of $\mathcal{K}(\bs)$ are deferred to Section \ref{sec:cone}.

\subsection{Analytical Solutions to UQP} \label{subsec:accurate}%(UQPs with known analytically accurate solution)
There exist cases for which the analytical global optima of UQP are easy to obtain. In this sub-section, we consider two such cases which will be used later to provide an approximate characterization of $\mathcal{K}(\bs)$.  A special example is the case in which $e^{j \arg(\bR)}$ (see the notation definition in \ref{subsec:organization}) is a rank-one matrix. More precisely, let 
$%\begin{eqnarray}
\bR=\bR_1 \odot (\widetilde{\bs} \widetilde{\bs}^H)
$ %\end{eqnarray}
 where $\bR_1$ is a real-valued Hermitian matrix with non-negative entries and $\widetilde{\bs} \in \Omega^n$ (a simple special case of this example is when $\bR$ is a rank-one matrix itself). In this case, it can be easily verified that $\bR_1 \in \mathcal{K}(\bone_{n \times 1})$. Therefore, using Theorem 1 one concludes that $\bR \in \mathcal{K}(\widetilde{\bs})$ i.e. $\bs=\widetilde{\bs}$ yields the global optimum of UQP. As another example, Theorem 2 considers the case for which several largest eigenvalues of the matrix $\bR$ are identical.
\begin{Theorem2} \label{th:uniqueness}
Let $\bR$ be a Hermitian matrix with eigenvalue decomposition $\bR=\bU \bSig \bU^H$. Suppose $\bSig$ is of the form
\begin{eqnarray}
\bSig=\mathbf{Diag}([\underbrace{\sigma_1~ \cdots ~\sigma_1}_{m~\mbox{times}}~ \sigma_2~\cdots~ \sigma_{n-m+1}]^T)
\end{eqnarray}
\begin{eqnarray}
\sigma_1 > \sigma_2 \geq \cdots \geq \sigma_{n-m+1} \nonumber
\end{eqnarray}
and let $\bU_m$ be the matrix made from the first $m$ columns of $\bU$. Now suppose $\widetilde{\bs} \in \Omega^n$ lies in the linear space spanned by the columns of $\bU_m$, i.e. there exists a vector $\balpha \in \complexC^m$ such that
\begin{eqnarray} \label{eq:sualpha}
\widetilde{\bs}=\bU_m \balpha.
\end{eqnarray}
Then $\widetilde{\bs}$ is a global optimizer of UQP.
\end{Theorem2}
\begin{proof}
Refer to the Appendix.
\end{proof}
\vspace{5pt}

%The problem of finding the global optimizer $\bs=\widetilde{\bs}$ of Theorem 2 can be approached by a cyclic minimization of the criterion
%\begin{eqnarray} \label{eq:cyclic}
%f( \bs,\balpha ) = \| \bs - \bU_m \balpha  \|_F 
%\end{eqnarray}
%where the solution is attained iff $f( \bs,\balpha )=0$. For any given $\balpha   \in \complexC^m$, the minimizing unimodular vector $\bs$ is given by
%$%\begin{eqnarray}
%\bs = e^{j \arg (\bU_m \balpha)}
%$. %\end{eqnarray}
%On the other hand, for given $\bs$ the minimizing $\balpha$ is obtained as $\bU_m^H \bs $. While minimizing (\ref{eq:cyclic}) is a non-convex problem as the UQP itself, it paves the way to a rather simple method for obtaining accurate solutions of UQP.

We end this section by noting that the solution to an UQP is not necessarily unique. For any set of unimodular vectors $ \{ \bs_1,\bs_2, \cdots , \bs_k \}$, $k \leq n$, we can use the Gram-Schmidt process to obtain a unitary matrix $\bU $ the first $k$ columns of which span the same linear space as $\bs_1,\bs_2, \cdots , \bs_k$. In this case, Theorem 2 suggests a method to construct a matrix $\bR$ (by choosing a $\bSig$ with $k$ identical largest eigenvalues) for which all  $\bs_1,\bs_2, \cdots , \bs_k$ are  global optimizers of the corresponding UQP.

\section{Specialized Local Optimization of UQP } \label{sec:local}
Due to its NP-hard nature, UQP has in general a highly multi-modal optimization objective. Finding and studying the local optima of UQP is not only useful to tackle the problem itself (particularly for UQP-related problems such as (\ref{eq:UQP-related})), but also to improve the UQP approximate solutions obtained by SDR or other optimization techniques. In this section, we introduce a computationally efficient procedure to obtain a local optimum of UQP.
% but with superior performance compared to any general local optimization method (as will be explained shortly). % We begin with the concept of hyper local optimization defined below.
%\begin{Definition1}
%Let $L_f$ represent the set of local optima of a function $f$. We say $H_f \subset L_f$ is a set of \textbf{hyper local optima} of $f$ if and only if $L_f$ can be partitioned  into the sets $\{ G_l \}_{l=1}^{|H_f|}$ with $s_l \in G_l$ such that
%\begin{eqnarray}
%\left\{ \begin{array}{ll}
%G_l \cap H_f = \{ s_l \} & \\
%f(s_l)>f(s) & \forall s \in G_l\backslash \{ s_l \}
%\end{array} \right.
%\end{eqnarray}
%Any $s_l \in H_f$ is a \textbf{hyper localhyper local optimum} of $f$.
%\end{Definition1}
%\vspace{4pt}
%Note that for any global optimum $\widetilde{s} \in \Omega^n$ of $f$, $\widetilde{s} \in H_f$ meaning that no global optimum can be excluded by a hyper local optimization scheme. 
%Next we begin with deriving a local optimization scheme for UQP. 

Note that, while the risk for this to happen in practice is nearly zero, local optimization methods can in theory converge to a saddle point. Consequently, in the sequel we let $L$ represent the set of all local optima and saddle points of UQP. Moreover, we assume that $\bR$ is positive definite. Consider the following relaxed version of UQP: 
\begin{eqnarray} \label{eq:UQP-relaxed}
\mbox{(RUQP) }~ \max_{\bs_1,\bs_2 \in \Omega^n} \Re (\bs_1^H \bR \bs_2)
\end{eqnarray}
We note that for fixed $\bs_2$ the maximizer of RUQP is given by 
\begin{eqnarray} \label{eq:s1}
\bs_1=e^{j \arg(\bR \bs_2)}.
\end{eqnarray}
Similarly, for any fixed $\bs_1$ the maximizer of RUQP is given by
\begin{eqnarray} \label{eq:s2}
\bs_2=e^{j \arg(\bR \bs_1)}.
\end{eqnarray}
In the following, we show that such a cyclic maximization of (\ref{eq:UQP-relaxed}) can be used to find local optima of UQP. It is not difficult to see that the criterion in (\ref{eq:UQP-relaxed}) increases and is upper bounded (by $\sum_{k,l} | \bR(k,l)|$) through the iterations in (\ref{eq:s1})-(\ref{eq:s2}), thus the said iterations are convergent in the sense of associated objective value. Next consider the identity
\begin{eqnarray} \label{eq:ident}
2 \Re (\bs_1^H \bR \bs_2) &=& \bs_1^H \bR \bs_1 + \bs_2^H \bR \bs_2 \\ \nonumber &-& (\bs_1-\bs_2)^H \bR (\bs_1-\bs_2).
\end{eqnarray}
%In case of $\bs_1 \neq \bs_2$, the positive semi-definiteness of $\bR$ implies that $(\bs_1-\bs_2)^H \bR (\bs_1-\bs_2)>0$. As a result,

Define $\varepsilon_\bs = \| \bs_1 - \bs_2\|_2^2$ and suppose that $\bs_2$ is fixed and its associated optimal $\bs_1$ is obtained by (\ref{eq:s1}).  It follows from (\ref{eq:ident}) that 
\begin{eqnarray} \label{eq:conv0}
\Re (\bs_1^H \bR \bs_2) \leq \frac{1}{2} \left(\bs_1^H \bR \bs_1 + \bs_2^H \bR \bs_2 \right) - \frac{\varepsilon_\bs}{2} \sigma_n (\bR).
\end{eqnarray}
Now suppose $\bs'_2$ is the optimal vector in $\Omega^n$ obtained by (\ref{eq:s2}) for the above $\bs_1$. Observe that $\Re (\bs_1^H \bR \bs'_2) \geq \Re (\bs_1^H \bR \bs_1)$ and that $\Re (\bs_1^H \bR \bs'_2) \geq \Re (\bs_1^H \bR \bs_2) \geq \Re (\bs_2^H \bR \bs_2) $ which imply
\begin{eqnarray} \label{eq:conv1}
\Re (\bs_1^H \bR \bs'_2) %&\geq& \max \{ \bs_1^H \bR \bs_1,  \bs_2^H \bR \bs_2 \} \\ \nonumber
&\geq& \frac{1}{2} \left(\bs_1^H \bR \bs_1 + \bs_2^H \bR \bs_2 \right) \\ \nonumber
&\geq& \Re (\bs_1^H \bR \bs_2) + \frac{\varepsilon_\bs}{2} \sigma_n (\bR).
\end{eqnarray}
It follows from (\ref{eq:conv1}) that
\begin{eqnarray} \label{eq:sandwich}
\varepsilon_\bs \leq \frac{2}{\sigma_n (\bR)} \left| \Re (\bs_1^H \bR \bs'_2) - \Re (\bs_1^H \bR \bs_2) \right|.
\end{eqnarray}
The right-hand side of (\ref{eq:sandwich}) vanishes through the cyclic minimization in (\ref{eq:s1})-(\ref{eq:s2}) which implies that $\varepsilon_\bs$ converges to zero at the same time. Note that the above arguments can be repeated for fixed $\bs_1$. We conclude that the iterations in (\ref{eq:s1})-(\ref{eq:s2}) are convergent and also that they cannot converge to $(\bs_1,\bs_2)$ with $\bs_1 \neq \bs_2 $. Moreover, as $\bs^H \bR \bs = \Re (\bs_1^H \bR \bs_2)$ for any $\bs_1=\bs_2=\bs$ then any local optimum $(\bs_1,\bs_2)$ of RUQP satisfying $\bs_1=\bs_2=\bs$ yields a local optimum $\bs$ of UQP. Based on the above discussions, the cyclic optimization of RUQP can be used to find local optima of UQP. Particularly, starting from any vector  $\bs^{(0)} \in \Omega^n$, the \emph{power method}-like iterations 
\begin{eqnarray} \label{eq:powermethod-like}
\bs^{(t+1)}= e^{j \arg(\bR \bs^{(t)})}
\end{eqnarray}
converge to an element in $L$. As an aside remark, we show that the objective of UQP is also increasing through the iterations of (\ref{eq:powermethod-like}). Using (\ref{eq:ident}) with $\bs_1=\bs^{(t+1)}$, and $\bs_2=\bs^{(t)}$ ($\bs^{(t+1)} \neq \bs^{(t)}$) implies that
\begin{eqnarray}
\bs^{(t+1) \, H } \bR \bs^{(t+1)} &>& - \bs^{(t) \, H} \bR \bs^{(t)} + 2 \Re (\bs^{(t+1) \, H} \bR \bs^{(t)}) \nonumber \\ 
 &\geq& \bs^{(t) \, H} \bR \bs^{(t)}.
\end{eqnarray}

Note that while (\ref{eq:powermethod-like}) can obtain the local optima of UQP, it might not converge to every of them. To observe this, let $\widetilde{\bs_1}$ be a local optimum of UQP and initialize (\ref{eq:powermethod-like}) with $\bs^{(0)}=\widetilde{\bs_1}$. Let $\widetilde{\bs_2}$ be another local optimum of UQP but with a larger value of UQP than that at $\widetilde{\bs_1}$. Now one can observe from (\ref{eq:ident}) that if $\widetilde{\bs_2}$ is sufficiently close to $\widetilde{\bs_1}$ then the above power method-like iterations can move away from $\widetilde{\bs_1}$, meaning that they can converge to another local optimum of UQP with a larger value of the UQP objective than that at $\widetilde{\bs_1}$. Therefore, (\ref{eq:powermethod-like}) bypasses some local optima of UQP with relatively small UQP objective values (which can be considered as an advantage compared to a general local optimization method). Moreover, one can note that there exist initializations for which (\ref{eq:powermethod-like}) leads to the global optimum of UQP (i.e. the global optimum is not excluded from the local optima to which (\ref{eq:powermethod-like}) can converge).

%In sum, (\ref{eq:powermethod-like}) has the following two properties:
%\begin{itemize}
%\item[a)] Starting from any initialization, the optimization procedure of (\ref{eq:powermethod-like}) either behaves like a local optimization method or yields a local optimum which has a better UQP objective than that obtained by a general local optimization method. This is achieved by bypassing some local optima with relatively small UQP objective values.
%\item[b)] There exist initializations for which (\ref{eq:powermethod-like}) leads to the global optimum of UQP (i.e. the global optimum is not excluded from the local optima to which (\ref{eq:powermethod-like}) can converge).
%\end{itemize} 

Next, we observe that any $\widetilde{\bs} \in L$ obtained by the above local optimization can be characterized by the equation
\begin{eqnarray} \label{eq:arg}
\arg(\widetilde{\bs})=\arg(\bR \widetilde{\bs}).
\end{eqnarray}
We refer to the subset of $L$ satisfying (\ref{eq:arg}) as the hyper points of UQP. Note that if $\widetilde{\bs} \in \Omega^n$ is a hyper point of UQP, then (\ref{eq:arg}) follows from the convergence of (\ref{eq:powermethod-like}). On the other hand, if  (\ref{eq:arg})  is satisfied, it implies the convergence of the iterations in (\ref{eq:powermethod-like}) and as a result $\widetilde{\bs}$ being a hyper point of UQP. The characterization given in (\ref{eq:arg}) is used below to motivate the characterization approach of Theorem 3.

\section{Results on the cone $\mathcal{K}(\bs)$}\label{sec:cone}
While a complete characterization of  $\mathcal{K}(\bs)$ cannot be expected (due to the NP-hardness of UQP), approximate characterizations of $\mathcal{K}(\bs)$ are possible. The goal of this section is to provide an approximate characterization of the cone $\mathcal{K}(\bs)$ which can be used to tackle the UQP problem. Our main result is as follows: %In particular, we derive a subset of $\mathcal{K}(s)$ that can be used to tackle the UQP
\begin{Theorem3} \label{th:cone-characterization}
For any given $\bs= ( e^{j \phi_1} ,  \cdots , e^{j \phi_n} )^T \in \Omega^n$, let $\{\bB_{k,l}\}$ be a set of matrices defined as
\begin{eqnarray}
\bB_{k,l} &=& (\be_k \be_l^H + \be_l \be_k^H) \odot (\bs \bs^H)
\end{eqnarray}
%or equivalently by their entries as
%\begin{eqnarray}
%\bB_{k,l} (k',l')= \left\{ \begin{array}{ll}
%e^{j (\phi_k - \phi_l)} & (k',l')=(k,l), \\
%0 & otherwise
%\end{array} \right.
%\end{eqnarray}
and  
%\begin{eqnarray}
$V_\bs=\{ \bB_{k,l}: 1 \leq k \leq l \leq n\} \cup \{- \bI_n \}$. 
%\end{eqnarray}
Let $\mathcal{C}(V_\bs)$ represent the convex cone associated with the basis matrices in $V_\bs$. Also let $\mathcal{C}_\bs$ represent the convex cone of matrices with $\bs$ being their dominant eigenvector (i.e the eigenvector corresponding to the maximal eigenvalue). Then for any $\bR \in \mathcal{K}(\bs)$, there exists $\alpha_0 \geq 0$ such that for all $\alpha \geq \alpha_0$,
\begin{eqnarray} \label{eq:cone_eq}
\bR+ \alpha \bs \bs^H \in  \mathcal{C}(V_\bs) \cup \mathcal{C}_\bs.
\end{eqnarray}
%In addition, for a positive definite matrix $\bR$ (\ref{eq:cone_eq}) holds only if $\bs$ is a local optimum of the UQP associated with $\bR$.
\end{Theorem3}
\vspace{5pt}
The proof of Theorem 3 will be presented in several steps (Theorems 4-7 and thereafter). Note that we show that (\ref{eq:cone_eq}) can be satisfied even if $\bs$ is a hyper point of UQP (satisfying (\ref{eq:arg})). However, since $\bs$ is the global optimum of UQP for all matrices in $\mathcal{C}_\bs$ and $\mathcal{C}(V_\bs)$, the case of $\alpha_0=0$ can occur only when $\bs$ is a global optimum of UQP associated with $\bR$. %[The main message to be delivered is that  $\mathcal{C}(V_\bs) \cup \mathcal{C}_\bs$ can be used as an approximation of $\mathcal{K}(s)$.]

Suppose $\bs$ is a hyper point of UQP associated with a given positive definite matrix $\bR$, and let $\theta_{k,l} = [\arg(\bR)]_{k,l}$. We define the matrix $\bR_+$ as
\begin{eqnarray}
\bR_+ (k,l) \! = \!  \left\{ \begin{array}{ll}
\! | \bR (k,l) |  \! \cos(\theta_{k,l}-(\phi_k - \phi_l)) \! & \!(k,l)\! \in \Theta, \\
0 & otherwise
\end{array} \right. 
\end{eqnarray}
where $\Theta$ represents the set of all $(k,l)$ such that $|\theta_{k,l}-(\phi_k - \phi_l)|< \pi / 2$. Now, let $\rho$ be a positive real number such that 
\begin{eqnarray}\label{eq:lambda_ineq}
\rho > \max_{(k,l) \notin  \Theta } \left\{ | \bR (k,l)\cos(\theta_{k,l}-(\phi_k - \phi_l)) | \right\}
\end{eqnarray}
and consider the sequence of matrices $\{\bR^{(t)} \}$ defined (in an iterative manner) by $\bR^{(0)}=\bR$, and
\begin{eqnarray}
\bR^{(t+1)}=\bR^{(t)} - (\bR_+^{(t)} - \rho \bone_{n \times n}) \odot (\bs \bs^H)
\end{eqnarray}
for $t\geq 0$. The next two theorems (whose proofs are given in the Appendix) study some useful properties of the sequence $\{\bR^{(t)} \}$.  %which pave the way for the results of Theorem 8 as an important step for the proof of Theorem 5.

\begin{Theorem4} 
$\{\bR^{(t)} \}$ is convergent in at most two iterations:
\begin{eqnarray}
\bR^{(t)} =  \bR^{(2)},~~\forall~ t \geq 2.
\end{eqnarray}
\end{Theorem4} 
%\vspace{3pt}
\begin{Theorem5}
$\bR^{(t)}$ is a function of $\rho$. Let $\rho$ and $\rho'$ both satisfy the criterion (\ref{eq:lambda_ineq}). At the convergence of $\{\bR^{(t)}\}$ (which is attained for $t=2$) we have:
\begin{eqnarray} \label{eq:lambda_diff}
\bR^{(2)} (\rho') = \bR^{(2)} (\rho) + (\rho'-\rho) (\bs \bs^H).
\end{eqnarray}
\end{Theorem5}
%\vspace{5pt}
Using the above results, Theorems 6 (whose proof is given in the Appendix) and 7 pave the way for a constructive proof of Theorem 3.
\begin{Theorem6}
If $\bs$ is a hyper point of the UQP associated with $\bR^{(0)}=\bR$ then it is also a hyper point of the UQPs associated with $\bR^{(1)} $ and $\bR^{(2)} $. Furthermore, $\bs$ is an eigenvector of $\bR^{(2)} $ corresponding to the eigenvalue $n \rho$.
\end{Theorem6}
%\vspace{3pt}
\begin{Theorem7}%\begin{Corollary1}
If $\bs$ is a hyper point of UQP for $\bR^{(0)}=\bR$ then it will be the dominant eigenvector of $\bR^{(2)} $ if $\rho$ is sufficiently large. In particular, let $\mu$ be the largest eigenvalue of $\bR^{(2)}$ which belongs to an eigenvector other than $\bs$. Then for any $\rho \geq \mu / n$, $\bs$ is a dominant eigenvector of $\bR^{(2)} $.
\end{Theorem7}
\begin{proof}
We know from Theorem 6 that $\bs$ is an eigenvector of $\bR^{(2)} $ corresponding to the eigenvalue $n \rho$. However, if $\bs$ is not the dominant eigenvector of  $\bR^{(2)} $, Theorem 5 implies that increasing $\rho$ would not change any of the eigenvalues/vectors of $\bR^{(2)} $ except that it increases the eigenvalue corresponding to $\bs$. As a result, for $\bs$ to be the dominant eigenvector of $\bR^{(2)}$ we only need $\rho$ to satisfy  $n \rho \geq \mu$ or equivalently $\rho \geq \mu / n$, which concludes the proof.
\end{proof}
\vspace{3pt}
Returning to Theorem 3, note that $\bR$ can be written as
\begin{eqnarray}
\bR&=& \bR^{(0)} \\ \nonumber &=& \bR^{(2)} + ( \bR_+^{(0)} +\bR_+^{(1)} ) \odot (\bs \bs^H) - 2 \rho \bs \bs^H.
\end{eqnarray}
For sufficiently large $\rho$ (satisfying both (\ref{eq:lambda_ineq}) and the condition of Theorem 7) we have that
\begin{eqnarray} \label{eq:lasteq}
\bR + 2 \rho \bs \bs^H =  \bR^{(2)} + ( \bR_+^{(0)} +\bR_+^{(1)} ) \odot (\bs \bs^H)
\end{eqnarray}
where $\bR^{(2)} \in \mathcal{C}_\bs$ and $(\bR_+^{(0)} +\bR_+^{(1)} ) \odot (\bs \bs^H) \in \mathcal{C}(V_\bs)$. Theorem 3 can thus be directly satisfied using Eq. (\ref{eq:lasteq}) with $\alpha_0= 2 \rho $. 

We conclude this section with two remarks. First of all, the above proof of Theorem 3 does not attempt to derive the minimal $\alpha_0$. In the following section we study a computational method to obtain an $\alpha_0$  which is as small as possible. Secondly, we can use $ \mathcal{C}(V_\bs) \cup \mathcal{C}_\bs$ as an approximate characterization of $\mathcal{K}(\bs)$ noting that the accuracy of such a characterization can be measured by the minimal value of $\alpha_0$. An explicit formulation of a sub-optimality guarantee for a solution of UQP based on the above $\mathcal{K}(\bs)$ approximation is derived in the following section.

\section{MERIT for UQP} \label{sec:global}

Using the previous results, namely the one-to-one mapping introduced in Theorem 1 and the approximation of  $\mathcal{K}(\bs)$ derived in Section \ref{sec:cone}, we build a sequence of matrices (for which the UQP global optima are known) whose distance from a given matrix is decreasing. The proposed iterative approach can be used to solve for the global optimum of UQP or at least to obtain a local optimum (with an upper bound on the sub-optimality of the solution). 
The sub-optimality guarantees are derived noting that the proposed method decreases an upper bound on the sub-optimality of the obtained UQP solution in each iteration.

We know from Theorem 3 that if $\bs$ is a hyper point of the UQP associated with $\bR$ then there exist matrices $\bQ_\bs \in \mathcal{C}_\bs$, $\bP_\bs \in \mathcal{C}(V_\bs)$ and a scalar $\alpha_0 \geq 0$ such that
\begin{eqnarray} \label{eq:decomposition}
\bR+ \alpha_0 \bs \bs^H = \bQ_\bs+ \bP_\bs.
\end{eqnarray}
Eq. (\ref{eq:decomposition}) can be rewritten as
\begin{eqnarray}
\bR+ \alpha_0 \bs \bs^H = (\bQ_\bone+ \bP_\bone) \odot (\bs \bs^H)
\end{eqnarray}
where $\bQ_\bone \in \mathcal{C}_\bone$, $\bP_\bone \in \mathcal{C}(V_\bone)$. 
We first consider the case  of $\alpha_0=0$ which corresponds to the global optimality of $\bs$. 

%\emph{Remark:} Depending on the difference between the values of the UQP objective at the first local optimum (i.e. the global optimum) and at the second local optimum, the efficiency of a computational method seeking the analytically global optimum of UQP might vary significantly. More precisely, if the UQP objective values at the first and the second local optima are sufficiently close, one might be able to reach an approximation guarantee which is virtually equal to $1$ but one cannot ensure that the obtained solution is a true analytically global optimum of UQP. Nevertheless, since the computational accuracy is limited anyway, a difference between an obtained virtually global optimum and an analytically global optimum of UQP is unavoidable.   \hfill $\blacksquare$
%\vspace{5pt}

\subsection{Global Optimization of UQP (the Case of $\alpha_0=0$)} \label{subsec:alpha00}
Consider the optimization problem:
\begin{eqnarray} \label{eq:opt_imp}
\min_{\bs \in \Omega^n, \bQ_\bone \in \mathcal{C}_\bone, \bP_\bone \in \mathcal{C}(V_\bone)} \| \bR - (\bQ_\bone+ \bP_\bone) \odot (\bs \bs^H) \|_F
\end{eqnarray}
Note that, as $\mathcal{C}_\bone \cup \mathcal{C}(V_\bone) $ is a convex cone, the global optimizers $\bQ_\bone $ and $\bP_\bone$ of (\ref{eq:opt_imp}) for any given $\bs$ can be easily found. On the other hand, the problem of finding an optimal $\bs$ for fixed $\bR_\bone =\bQ_\bone+ \bP_\bone$ is non-convex and hence more difficult to solve globally (see below for details).

We will assume that $\bR_\bone$ is a positive definite matrix. To justify this  assumption let $\overline{\bR}=\bR \odot (\bs \bs^H)^* $ and note that the eigenvalues of $\overline{\bR}$ are exactly the same as those of $\bR$, hence $\overline{\bR}$ is positive definite. Suppose that we have
\begin{eqnarray} \label{eq:cond}
 \left\{ \begin{array}{l}
 \bx^H \overline{\bR}\bx> \varepsilon, ~~ \forall \mbox{ unit-norm } \bx \in \complexC^{n \times 1}\\
\| \overline{\bR} - \bR_\bone \|_F \leq \varepsilon
\end{array} \right. 
\end{eqnarray}
for some $\varepsilon \geq 0$. It follows from (\ref{eq:cond}) that
\begin{eqnarray}
 \bx^H \bR_\bone  \bx &\geq&  \bx^H \overline{\bR} \bx - |  \bx^H \overline{\bR} \bx -  \bx^H \bR_\bone  \bx | \\ \nonumber
 &>& \varepsilon - | \bx^H (\overline{\bR} - \bR_\bone ) \bx | \\ \nonumber
 &\geq& \varepsilon - |\sigma_1(\overline{\bR} - \bR_\bone) | \\ \nonumber
 &\geq& \varepsilon - \| \overline{\bR} - \bR_\bone \|_F  \geq 0
\end{eqnarray}
which implies that $\bR_\bone$ is also a positive definite matrix. The conditions in (\ref{eq:cond}) can be met as follows. By considering only the component of $\bR_\bone$ in $\mathcal{C}(V_\bone)$ (namely $\bP_\bone$) we observe that any positive (i.e. with $\lambda>0$) diagonal loading of $\bR$, which leads to the same diagonal loading of $\overline{\bR}$ (as $\overline{\bR} + \lambda \bI = \bR \odot (\bs \bs^H)^* + \lambda \bI = (\bR+ \lambda \bI ) \odot (\bs \bs^H)^* $), will be absorbed in $\bP_\bone$. Therefore, a positive diagonal loading of $\bR$ does not change $\| \overline{\bR} - \bR_\bone \|_F$ but increases $\bx^H \overline{\bR} \bx$ by $\lambda$. We also note that due to $\| \overline{\bR} - \bR_\bone \|_F$ being monotonically decreasing through the iterations of the method, if the conditions in (\ref{eq:cond}) hold for the solution obtained in any iteration, it will hold for all the iterations afterward. 
%However, based on empirical observations, in order to satisfy (\ref{eq:cond}) it is not strictly necessary to apply such diagonal loading of $\bR$.

In the following, we study a suitable diagonal loading of $\bR$ that ensures meeting the conditions in (\ref{eq:cond}). Next the optimization of the function in (\ref{eq:opt_imp}) is discussed through a separate optimization over the three variables of the problem.  \\
$\bullet$~ \emph{Diagonal loading of $\bR$:} As will be explained later, we can compute $\bQ_\bone$ and $\bP_\bone$, (hence $\bR_\bone=\bQ_\bone+\bP_\bone$) for any initialization of $\bs$. In order to guarantee the positive definiteness of $\bR_\bone$, define
\begin{eqnarray} \label{eq:dl1}
\varepsilon_0 \triangleq \| \overline{\bR} - \bR_\bone \|_F.
\end{eqnarray}
Then we suggest to diagonally load $\bR$ with $\lambda > \lambda_0 = -\sigma_n(\bR) + \varepsilon_0$:
\begin{eqnarray} \label{eq:dl2}
\bR \leftarrow \bR + \lambda \bI.
\end{eqnarray}
$\bullet$~ \emph{Optimization with respect to $\bQ_\bone$:} We restate the objective function of (\ref{eq:opt_imp}) as
\begin{eqnarray}
&~& \| \bR - (\bQ_\bone+ \bP_\bone) \odot (\bs \bs^H) \|_F \\ \nonumber
&=& \| \underbrace{\left( \bR \odot (\bs \bs^{H})^* -  \bP_\bone \right)}_{\bR_Q} - \bQ_\bone \|_F.
\end{eqnarray}
Given $\bR_Q$, (\ref{eq:opt_imp}) can be written as
\begin{eqnarray} \label{eq:opt_q}
\min_{\bQ_\bone \in \mathcal{C}_\bone} \| \bR_Q - \bQ_\bone \|_F.
\end{eqnarray}
In \cite{nearest-doubly-stochastic}, the authors have derived an explicit solution for the optimization problem
\begin{eqnarray} \label{eq:exp_sol}
\min_{\bQ_\bone} \| \bR_Q - \bQ_\bone \|_F ~~~~\\ \nonumber
\mbox{s.t. } \bQ_\bone \bone = \rho \bone. \mbox{ ($\rho=$given)}
\end{eqnarray}
The explicit solution of (\ref{eq:exp_sol}) is given by
\begin{eqnarray} \label{eq:frob_imply}
\bQ_\bone (\rho) &=& \rho \bI_n %\\ \nonumber &+& 
+(\bI_n - \frac{\bone_{n \times n}}{n}) (\bR_Q -  \rho \bI_n) (\bI_n - \frac{\bone_{n \times n}}{n})  \\ \nonumber
&=&  \bR_Q +  \frac{\rho}{n} \bone_{n \times n} %\\ \nonumber &-& 
-\frac{2}{n} (\bR_Q \bone_{n \times n}) + \frac{1}{n^2} (\bone_{n \times n}  \bR_Q \bone_{n \times n} )
\end{eqnarray}
Note that 
\begin{eqnarray}
\bQ_\bone (\rho') - \bQ_\bone (\rho)= (\rho'-\rho) (\bone_{n \times 1}/\sqrt{n}) (\bone_{n \times 1}/\sqrt{n})^T %\nonumber 
\end{eqnarray}
%\begin{eqnarray}
%\end{eqnarray}
which implies that except for the eigenpair $(\bone_{n \times 1}/\sqrt{n},\rho)$, all other eigenvalue/vectors are independent of $\rho$. Let $\rho_0$ represent the maximal eigenvalue of $\bQ_\bone (0)$ corresponding to an eigenvector other than $\bone_{n \times 1}/\sqrt{n}$. Therefore, (\ref{eq:opt_q}) is equivalent to
\begin{eqnarray} \label{eq:boils}
 \min_{\rho}  \| \bR_Q - \bQ_\bone (\rho) \|_F \\ \nonumber
 \mbox{s.t. }  \rho \geq \rho_0.~~~~~~~
\end{eqnarray}
It follows from (\ref{eq:frob_imply})  that
\begin{eqnarray} \label{eq:75}
\| \bR_Q - \bQ_\bone (\rho) \|^2_F =  \sum_{k=1}^{n} n \left| \frac{\rho}{n}-  \frac{2 G_k}{n} + \frac{H}{n^2} \right|^2
\end{eqnarray}
where $G_k$ and $H$ are the sum of the $k^{th}$ row and, respectively, the sum of all entries of $\bR_Q $. The $\rho$ that minimizes (\ref{eq:75}) is given by
\begin{eqnarray}
\rho = \frac{1}{n} \sum_{k=1}^{n} \Re \left(2 G_k - \frac{H}{n} \right) = \frac{H}{n}
\end{eqnarray}
which implies that the minimizer $\rho=\rho_\star$ of (\ref{eq:boils}) is equal to
\begin{eqnarray}
\rho_\star = \left\{ \begin{array}{ll}
 \frac{H}{n} & \frac{H}{n} \geq \rho_0, \\
 \rho_0 & \mbox{otherwise.}
\end{array} \right.
\end{eqnarray}
Finally, the optimal solution $\bQ_\bone$ to (\ref{eq:opt_q}) is given by
\begin{eqnarray} \label{eq:bq1_opt}
\bQ_\bone = \bQ_\bone (\rho_\star).
\end{eqnarray} 
$\bullet$~ \emph{Optimization with respect to $\bP_\bone$:} Similar to the previous case, (\ref{eq:opt_imp}) can be rephrased as
\begin{eqnarray} \label{eq:opt_p}
\min_{\bQ_\bone \in \mathcal{C}(V_\bone)} \| \bR_P - \bP_\bone \|_F
\end{eqnarray}
where $\bR_P = \bR \odot (\bs \bs^H)^* -  \bQ_\bone$.  The solution of (\ref{eq:opt_p}) is simply given by
\begin{eqnarray} \label{eq:bp1_opt}
\bP_\bone (k,l)= \left\{ \begin{array}{ll}
 \bR'_P (k,l) & \bR'_P(k,l) \geq 0 \mbox{ or } k=l, \\
 0 & \mbox{otherwise}
\end{array} \right.
\end{eqnarray}
where $\bR'_P = \Re\{\bR_P\}$.\\
$\bullet$~ \emph{Optimization with respect to $\bs$:} Suppose that $\bQ_\bone $ and $\bP_\bone$ are given and that $\bR_\bone=\bQ_\bone+\bP_\bone$ is a positive definite matrix (see the discussion on this aspect following Eq. (\ref{eq:opt_imp})). We consider a relaxed version of (\ref{eq:opt_imp}), %n apparently
\begin{eqnarray} \label{eq:opt_imp_relaxed}
\min_{\bs_1,\bs_2 \in \Omega^n } \| \bR -  \bR_\bone \odot (\bs_1 \bs_2^H) \|_F 
\end{eqnarray}
The objective function in (\ref{eq:opt_imp_relaxed}) can be re-written as
\begin{eqnarray} \label{eq:s1s2}
&~& \| \bR -  \bR_\bone \odot (\bs_1 \bs_2^H) \|^2_F   \\ \nonumber
&=& \| \bR -  \mathbf{Diag}(\bs_1) \,  \bR_\bone  \, \mathbf{Diag}( \bs_2^*) \|^2_F   \\ \nonumber
&=& \mbox{tr}(\bR^2) + \mbox{tr}( \bR_\bone^2) %\\ \nonumber &-& 
- 2 \Re \{\mbox{tr}(\bR ~\mathbf{Diag}(\bs_1) \, \bR_\bone  \, \mathbf{Diag}( \bs_2^*))\}.
\end{eqnarray}
Note that only the third term of (\ref{eq:s1s2}) is a function of $\bs_1$ and $\bs_2$. Moreover, it can be verified that \cite{horn1990matrix}
\begin{eqnarray}
\mbox{tr}(\bR ~\mathbf{Diag}(\bs_1) \, \bR_\bone  \, \mathbf{Diag}( \bs_2^*)) = \bs_2^H (\bR \odot \bR_\bone^T) \bs_1.
\end{eqnarray}
As $\bR \odot \bR_\bone^T$ is positive definite, we can employ the power method-like iterations introduced in (\ref{eq:powermethod-like}) to obtain a solution to (\ref{eq:opt_imp}) i.e. starting from the current $\bs=\bs^{(0)}$, a local optimum of the problem can be obtained by the iterations
\begin{eqnarray} \label{eq:respect_s}
\bs^{(t+1)}= e^{j \arg((\bR \odot \bR_\bone^T) \bs^{(t)})}.
\end{eqnarray}

Finally, the proposed algorithmic optimization of (\ref{eq:opt_imp}) based on the above results is summarized in Table~\ref{table:alpha_0_0}-A.

\begin{table}[tp]
\caption{The MERIT Algorithm} \label{table:alpha_0_0} \centering
\begin{tabular}{p{4.0in}}%{p{3.3in}}
\hline 
(A) The case of $\alpha_0=0$\\ \hline \vspace{1.5pt}
\textbf{Step 0}: Initialize the variables $\bQ_\bone$ and $\bP_\bone$ with $\bI$. Let $\bs$ be a random vector in $ \Omega^n$. \\
\textbf{Step 1}: Perform the diagonal loading of $\bR$ as in (\ref{eq:dl1})-(\ref{eq:dl2}) (note that this diagonal loading is sufficient to keep $\bR_\bone=\bQ_\bone+\bP_\bone$ always positive definite).\\
\textbf{Step 2}: Obtain the minimum of (\ref{eq:opt_imp}) with respect to $\bQ_\bone$ as in (\ref{eq:bq1_opt}).  \\
\textbf{Step 3}: Obtain the minimum of (\ref{eq:opt_imp}) with respect to $\bP_\bone$ using (\ref{eq:bp1_opt}). \\
\textbf{Step 4}: Minimize (\ref{eq:opt_imp}) with respect to $\bs$ using (\ref{eq:respect_s}). \\
\textbf{Step 5}: Goto step 2 until a stop criterion is satisfied, e.g. $\| \bR - (\bQ_\bone+ \bP_\bone) \odot (\bs \bs^H) \|_F \leq \epsilon_0$ (or if the number of iterations exceeded a predefined maximum number).
\vspace{7pt}\\
\hline
(B) The case of $\alpha_0>0$ \\
 \hline \vspace{1.5pt}
\textbf{Step 0}: Initialize the variables $(\bs, \bQ_\bone ,\bP_\bone) $ using the results obtained by the optimization of (\ref{eq:opt_imp}) as in Table~\ref{table:alpha_0_0}-A. \\
\textbf{Step 1}: Set $\delta$ (the step size for increasing $\alpha_0$ in each iteration). Let $\delta_0$ be the minimal  $\delta$ to be considered and $\alpha_0=0$. \\
\textbf{Step 2}: Let $\alpha_0^{pre}=\alpha_0$, $\alpha_0^{new}=\alpha_0+\delta$ and $\bR' = \bR + \alpha_0^{new} \bs \bs^H$. \\
\textbf{Step 3}: Solve (\ref{eq:opt_imp2}) using the steps 2-5 in Table~\ref{table:alpha_0_0}-A (particularly step 4 must be applied to (\ref{eq:local_s})).\\
\textbf{Step 4}: If $\| \bR' - (\bQ_\bone+ \bP_\bone) \odot (\bs \bs^H) \|_F \leq \epsilon_0$ do:\\ 
\begin{itemize}
\item \textbf{Step 4-1}: If $\delta \geq \delta_0$, let $\delta \leftarrow \delta/2$ and initialize  (\ref{eq:opt_imp2}) with the previously obtained variables $(\bs, \bQ_\bone ,\bP_\bone) $  for $\alpha_0=\alpha_0^{pre}$. Goto step 2.
\item \textbf{Step 4-2}: If $\delta < \delta_0$, stop.
\end{itemize}
Else, let $\alpha_0=\alpha_0^{new}$ and goto step 2.\vspace{8pt}\\
 \hline
\end{tabular}
\end{table}

\subsection{Achieving a Local Optimum of UQP (the Case of $\alpha_0>0$)} \label{subsec:alpha0p}
There exist examples for which the objective function in (\ref{eq:opt_imp}) does not converge to zero. As a result, the proposed method cannot obtain a global optimum of UQP in such cases. However, it is still possible to obtain a local optimum of UQP for some $\alpha_0>0$. To do so, we solve the optimization problem,
\begin{eqnarray} \label{eq:opt_imp2}
\min_{\bs \in \Omega, \bQ_\bone \in \mathcal{C}_\bone, \bP_\bone \in \mathcal{C}(V_\bone)} \| \bR' - (\bQ_\bone+ \bP_\bone) \odot (\bs \bs^H) \|_F
\end{eqnarray}
with $\bR' = \bR + \alpha_0 \bs \bs^H$, for increasing $\alpha_0$. The above optimization problem can be tackled using the same tools as proposed for (\ref{eq:opt_imp}). In particular, note that increasing $\alpha_0$ decreases  (\ref{eq:opt_imp2}). To observe this, suppose that the solution $(\bs,\bQ_\bone,\bP_\bone)$ of (\ref{eq:opt_imp2}) is given for an $\alpha_0 \geq 0$. The minimization of (\ref{eq:opt_imp2}) with respect to $\bQ_\bone$ for $\alpha_0^{new}=\alpha_0+ \delta$ ($\delta>0$) yields $\widetilde{\bQ}_\bone \in \mathcal{C}_\bone$ such that
\begin{eqnarray}
&& \| \bR+ \alpha_0^{new} \bs \bs^H  - (\widetilde{\bQ}_\bone+ \bP_\bone) \odot (\bs \bs^H) \|_F \\ \nonumber
&\leq&  \| \bR+  \alpha_0^{new} \bs \bs^H  - ((\bQ_\bone+ \delta \bone \bone^T)+ 
\bP_\bone) \odot (\bs \bs^H) \|_F \\ \nonumber
&=&  \| \bR+ \alpha_0 \bs \bs^H  - (\bQ_\bone+ \bP_\bone) \odot (\bs \bs^H) \|_F
\end{eqnarray}
where $\bQ_\bone+ \delta \bone \bone^T \in \mathcal{C}_\bone$. The optimization of (\ref{eq:opt_imp2}) with respect to $\bP_\bone$ can be dealt with as before (see (\ref{eq:opt_imp}) and it leads to a further decrease of the objective function. Furthermore, 
\begin{eqnarray} \label{eq:local_s}
&& \| \bR+ \alpha_0 \bs \bs^H  - (\bQ_\bone+ \bP_\bone) \odot (\bs \bs^H) \|_F \\ \nonumber
&=&  \| \bR+ \lambda' \bI  - (\bQ_\bone +\bP_\bone - \alpha_0 \bone \bone^T + \lambda' \bI ) \odot (\bs \bs^H) \|_F
\end{eqnarray}
which implies that a solution $\bs$ of (\ref{eq:opt_imp2}) can be obtained via optimizing (\ref{eq:local_s}) with respect to $\bs$   in a similar way as we described for (\ref{eq:opt_imp}) provided that $\lambda' \geq 0$ is such that $\bQ_\bone +\bP_\bone - \alpha_0 \bone \bone^T + \lambda' \bI $ is positive definite.
%as the additional term in $\bR'$ will be absorbed in $\bQ_\bone$ and $\bP_\bone$ at the first iteration (after including the additional term). Therefore, the optimization of (\ref{eq:opt_imp2}) can be viewed as the secondary step of the optimization of (\ref{eq:opt_imp}) when the convergence is to a value greater than zero. 
Finally, note that the obtained solution $(\bs, \bQ_\bone ,\bP_\bone) $ of  (\ref{eq:opt_imp}) can be used to initialize the corresponding variables in  (\ref{eq:opt_imp2}). In effect, the solution of (\ref{eq:opt_imp2}) for any $\alpha_0$ can be used for the initialization of (\ref{eq:opt_imp2}) with an increased $\alpha_0$.

Based on the above discussion and the fact that small values of $\alpha_0$ are of interest, a bisection approach can be used to obtain $\alpha_0$. The proposed method for obtaining a local optimum of UQP along with the corresponding $\alpha_0$ is described in Table~\ref{table:alpha_0_0}-B.

\subsection{Sub-optimality Analysis} \label{subsec:subopt}
%We draw the attention of the reader to the fact that for any UQP optimization scheme, the obtained UQP solution can be better than the provided sub-optimality guarantee. 
In this sub-section, we show that the proposed method can provide a sub-optimality guarantee ($\gamma$) that is close to $1$. Let $\alpha_0=0$ (as a result $\bR'=\bR$) and define
\begin{eqnarray}
\bE   \triangleq  \bR' - \underbrace{(\bQ_\bone+ \bP_\bone) \odot (\bs \bs^H)}_{\bR_\bs}
\end{eqnarray}
 where $\bQ_\bone \in \mathcal{C}_\bone$ and $\bP_\bone \in \mathcal{C}(V_\bone)$. The global optimum of the UQP associated with $\bR_\bs$ is $\bs$. We have that
\begin{eqnarray} \label{eq:up-bound}
\max_{\bs' \in \Omega^n} \bs'^H \bR \bs' &\leq& \max_{\bs' \in \Omega^n} \bs'^H \bR_\bs \bs' + \max_{\bs' \in \Omega^n} \bs'^H \bE \bs' 
\\ \nonumber
&\leq & \max_{\bs' \in \Omega^n} \bs'^H \bR_\bs \bs' + n \sigma_1(\bE)
\\ \nonumber
&=&  \bs^H \bR_\bs \bs + n \sigma_1(\bE)
\end{eqnarray}
Furthermore, 
\begin{eqnarray} \label{eq:low-bound}
\max_{\bs' \in \Omega^n} \bs'^H \bR \bs' &\geq& \max_{\bs' \in \Omega^n} \bs'^H \bR_\bs \bs' + \min_{\bs' \in \Omega^n} \bs'^H \bE \bs' 
\\ \nonumber
&\geq & \max_{\bs' \in \Omega^n} \bs'^H \bR_\bs \bs' + n \sigma_n(\bE)
\\ \nonumber
&= & \bs^H \bR_\bs \bs + n \sigma_n(\bE)
\end{eqnarray}
As a result, an upper bound and a lower bound on the objective function for the global optimum of (\ref{eq:opt_imp}) can be obtained \emph{at each iteration}. Furthermore, as
\begin{eqnarray}
| \sigma_1(\bE)| \leq  \| \bE \|_F, ~| \sigma_n(\bE)| \leq  \| \bE \|_F
\end{eqnarray}
if $\| \bE \|_F $ converges to zero %(which is the case in many observed examples), 
we conclude for (\ref{eq:up-bound}) and (\ref{eq:low-bound}) that
\begin{eqnarray}
\max_{\bs' \in \Omega^n} \bs'^H \bR \bs'  = \bs^H \bR_\bs \bs = \bs^H \bR \bs %\max_{\bs' \in \Omega} \bs'^H \bR_\bs \bs'
\end{eqnarray}
and hence $\bs$ is the global optimum of the UQP associated with $\bR$ (i.e. a sub-optimality guarantee of $\gamma=1$ is achieved).

Next, suppose that we have to increase $\alpha_0$ in order to obtain the convergence of  $\| \bE \|_F $ to zero. In such a case, we have that
$%\begin{eqnarray}
\bR = \bR_\bs - \alpha_0 \bs \bs^H
$ %\end{eqnarray}
and as a result, 
%\begin{eqnarray}
$\max_{\bs' \in \Omega^n} \bs'^H \bR_\bs \bs' - \alpha_0 n^2 \leq \ \max_{\bs' \in \Omega^n} \bs'^H \bR \bs' \leq \max_{\bs' \in \Omega^n} \bs'^H \bR_\bs \bs'$ 
%\end{eqnarray}
or equivalently, 
\begin{eqnarray}
\bs^H \bR_\bs \bs - \alpha_0 n^2 \leq \max_{\bs' \in \Omega^n} \bs'^H \bR \bs' \leq  \bs^H \bR_\bs \bs. 
\end{eqnarray}
The provided sub-optimality guarantee is thus given by
\begin{eqnarray} \label{eq:subopt}
\gamma= \frac{\bs^H \bR \bs}{\bs^H \bR_\bs \bs} = 1- \frac{\alpha_0 n^2}{\bs^H \bR_\bs \bs}. 
\end{eqnarray}

Note that while solving the optimization problem (\ref{eq:opt_imp2}) does not necessarily yield the exact optimal solution to UQP, the so-obtained solution can be still optimal. We also note that (\ref{eq:subopt}) generally yields tighter sub-optimality guarantees than the currently known approximation guarantee (i.e. $\pi /4$ for SDR). The following section provides  empirical evidence for such a fact.

\section{Numerical Examples}\label{sec:numerical}
In order to examine the performance of the proposed method, several numerical examples will be presented. Random Hermitian matrices $\bR$ are generated using the formula
\begin{eqnarray} \label{eq:xdef}
\bR = \sum_{k=1}^{n} \bx_k \bx_k^H
\end{eqnarray}
where $\{\bx_k\}$ are random vectors in $\complexC^n$ whose real-part and imaginary-part elements are i.i.d. with a standard Gaussian distribution $\mathcal{N}(0,1)$. In all cases, we stopped the iterations when $\| \bE \|_F \leq 10^{-9}$. 

We use the MERIT algorithm to solve the UQP for a random positive definite  matrix of size $n=16$. The obtained values of the UQP objective for the true matrix ($\bR$) and the approximated matrix ($\bR_\bs$) as well as the sub-optimality bounds (derived in (\ref{eq:up-bound}) and (\ref{eq:low-bound})) are depicted in Fig. 1 versus the iteration number. In this example, a sub-optimality guarantee of $\gamma=1$ is achieved which implies that the method has successfully obtained the global optimum of the considered UQP. A computational time of 3.653 sec was required on a standard PC to accomplish the task.

\begin{figure*}[tp] 
\centering
\subfigure[]{\includegraphics[width=8.5cm]{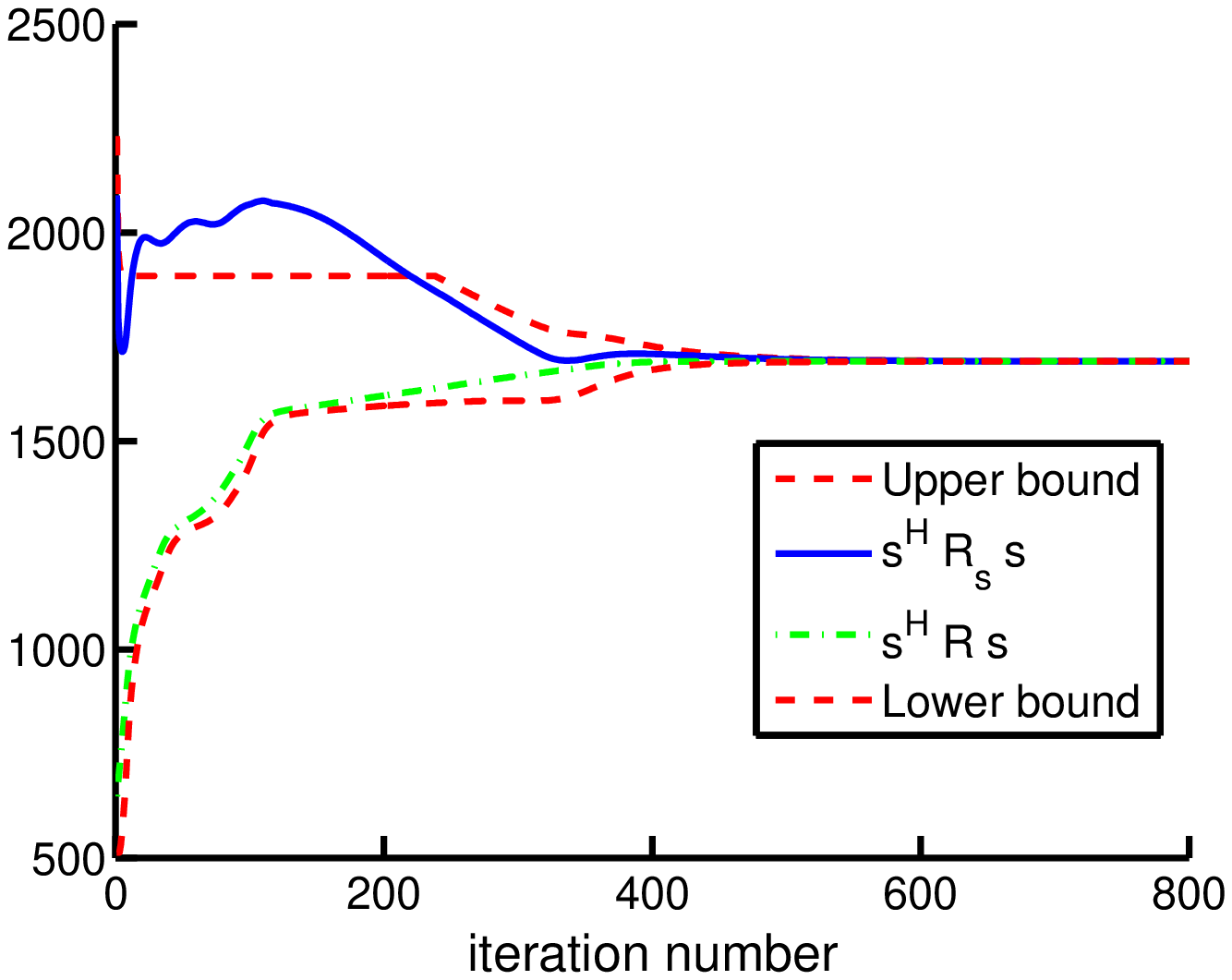}\label{fig:rgb}}\subfigure[]{\includegraphics[width=8.5cm]{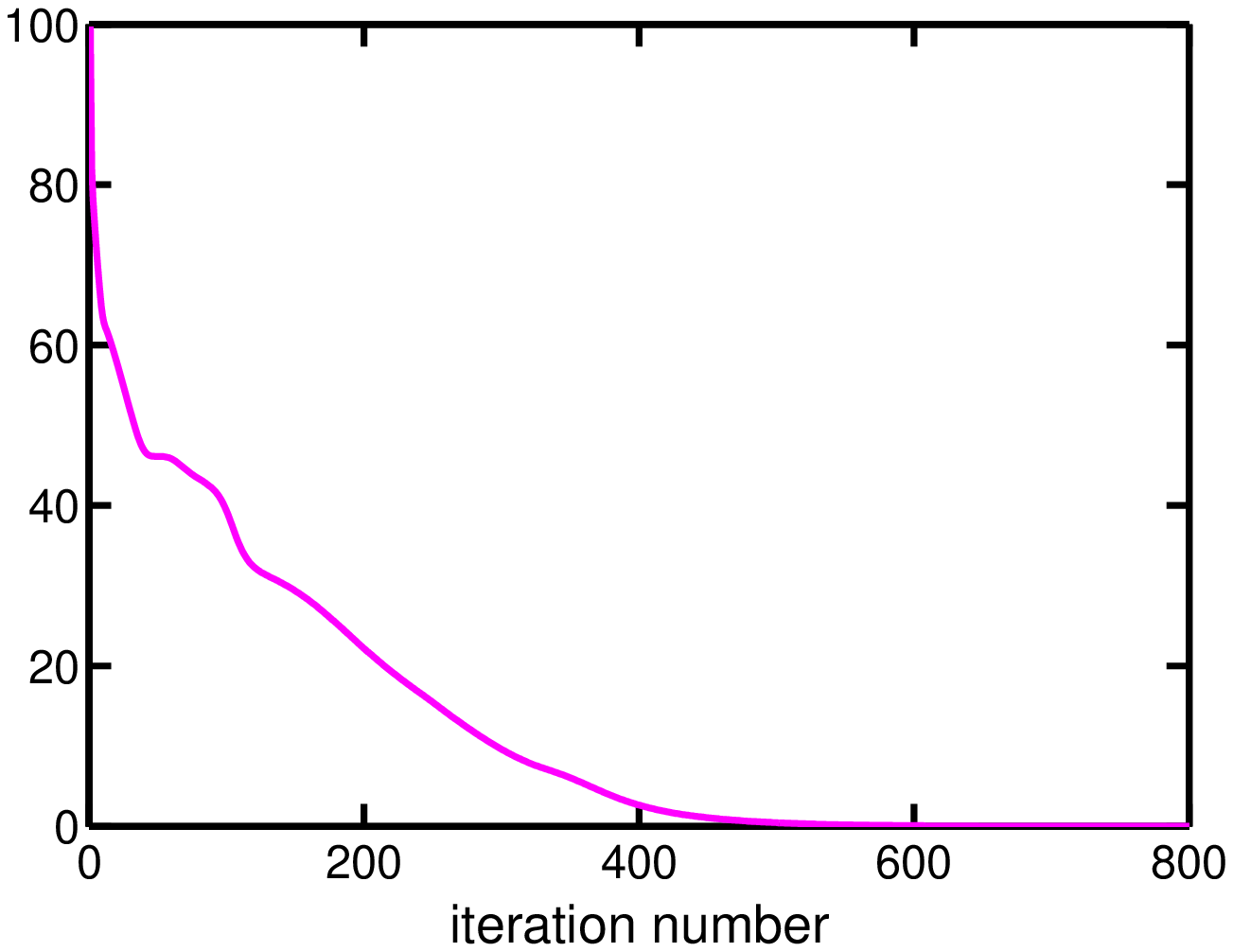}\label{fig:obj2}}
\caption{Different metrics versus the iteration number for an UQP solved by MERIT. (a) the UQP objective corresponding to the true matrix ($\bR$), the approximated matrix ($\bR_\bs$) and also the upper/lower bounds at each iteration. The sub-optimality bounds are updated using (\ref{eq:up-bound})-(\ref{eq:low-bound}). (b) the criterion $\| \bE \|_F=\|\bR - \bR_s \|_F$ (it reaches values which are practically zero).}
\end{figure*}

Next, we solve the UQP for $20$ full-rank random positive definite matrices of sizes $n \in \{ 8,16,32,64 \}$. Inspired by \cite{low-rank} and \cite{rankdef_binary}, we also consider rank-deficient matrices 
%\begin{eqnarray}
$\bR = \sum_{k=1}^{d} \bx_k \bx_k^H$ 
%\end{eqnarray}
where $\{\bx_k\}$ are as in (\ref{eq:xdef}), but $d \ll n$. The performance of MERIT for different values of $d$ is shown in Table~\ref{table_rank_def}.  Interestingly, the solution of UQP for rank-deficient matrices appears to be obtained more efficiently than for the full-rank matrices. For each problem solved by MERIT, we also let the SDR algorithm of \cite{DeMaio-PAR} use the same computational time for solving the problem. The SDR algorithm is able to solve the problem only if its core semi-definite program can be solved within the available time. Any remaining time is dedicated to the randomization procedure. %The comparison of UQP objective values is accomplished for cases in which SDR can solve the UQP which thus is likely to favorize SDR. 
The results can be found in Table~\ref{table_rank_def}. Note that the maximum UQP objective values obtained by MERIT and SDR were nearly identical in those cases in which SDR was able to solve the UQPs in the same amount of time as MERIT. Note also that given the solutions obtained by MERIT and SDR as well as the sub-optimality guarantee of MERIT, a case-dependent sub-optimality guarantee for SDR can be computed as
\begin{eqnarray}
\gamma_{\mbox{SDR}} \triangleq \gamma_{\mbox{MERIT}} \left(\frac{v_{SDR}}{v_{MERIT}} \right).
\end{eqnarray}
This can be used to examine the goodness of the solutions obtained by SDR. 

\begin{table*}
\begin{center}
\footnotesize
\begin{tabular}{|c|c|p{2.0cm}|p{2.0cm}|p{2.0cm}|p{2.0cm}|p{2.0cm}|}
  \hline
  $n$ & Rank ($d$) & \#{problems for which $\gamma=1$} & {Average $\gamma$} & {Minimum $\gamma$} & {Average CPU time (sec)} & \#{problems solved by SDR} \\ %& $\mathbb{E} \left[ \frac{v_{SDR}}{v_{MERIT}} \right]$\\

  \hline 
  \hline 
   \textbf{8} & \textit{2} & $17$& $0.9841$& $0.8184$& $0.13$ & $4$ \\ \cline{2-7}
   & \textit{8}& $16$& $0.9912$& $0.9117$& $0.69$ & $7$ \\ 
  \hline \hline
    & \textit{2} & $15$& $0.9789$& $0.8301$& $1.06$ & $2$ \\ \cline{2-7}
 	\textbf{16} & \textit{4} & $13$& $0.9773$& $0.8692$& $1.58$ & $10$ \\ \cline{2-7}
	 & \textit{16}& $4$& $0.9610$& $0.8693$& $3.54$ & $13$\\ 
\hline \hline
     & \textit{2} & $9$& $0.9536$& $0.8190$& $47.04$ & $3$ \\ \cline{2-7}
 	\textbf{32} & \textit{6} &$4$& $0.9077$& $0.8106$& $55.59$ & $7$ \\ \cline{2-7}
	 & \textit{32} & $2$& $0.9031$& $0.8021$& $94.90$ & $16$ \\
\hline \hline
    & \textit{2} &$3$& $0.8893$& $0.8177$& $406.56$ & $4$ \\ \cline{2-7}
 	\textbf{64} & \textit{8} & $1$& $0.8567$& $0.7727$& $560.35$& $10$ \\ \cline{2-7}
	 & \textit{64} & $0$& $0.8369$& $0.7811$& $1017.69$& $15$  \\	
  \hline
\end{tabular}
\end{center}
\caption{Comparison of the performance of MERIT (see Table~\ref{table:alpha_0_0}) and SDR \cite{DeMaio-PAR} when solving the UQP for $20$ random positive definite matrices of different sizes $n$ and ranks $d$.}
\label{table_rank_def}
\end{table*}

Besides random matrices, we also consider several other matrix structures for which solving the UQP using the proposed method is not ``hard", as explained below. %The provided examples are of more practical significance compared to the rank-deficient matrices. Consider the disturbance matrices in the following.
\begin{itemize}
\item  \emph{Case 1:} An exponentially shaped disturbance matrix \cite{DeMaio-similarity} with correlation coefficient $\eta = 0.8$,
\begin{eqnarray}
\bM(k,l)=\eta^{|k-l|},~~~1\leq k,l \leq n .
\end{eqnarray}
\item  \emph{Case 2:} A disturbance matrix with the structure
\begin{eqnarray}
\bM(k,l)=\eta_1^{|k-l|} e^{j2\pi \rho (k-l)} + 10  \eta_2 ^{|k-l|} + 10^{-2} \bI(k,l),~~~1\leq k,l \leq n 
\end{eqnarray}
whose terms represent the effects of sea clutter, land clutter and thermal noise, respectively. The values of $(\eta_1,\eta_2,\rho)$ are set to $(0.8,0.9,0.2)$ in accordance to an example provided in \cite{Demaio-maxmin}.
\item  \emph{Case 3:} A disturbance matrix accounting for both discrete clutter scatterers and thermal noise \cite{DeMaio-PAR},
\begin{eqnarray}
\bM= \sum_{k=1}^{n_c} \eta_k  \bp_{v_{d,k}} \bp_{v_{d,k}}^H + \eta \bI
\end{eqnarray}
where $n_c=10$, $\eta_k=10^3$, $v_{d,k}=(k-1)/2$, 
\begin{eqnarray} \label{eq:p_struct}
\bp_{v_{d,k}}=( 1, e^{j2\pi v_{d,k}}, \cdots, e^{j2\pi (n-1) v_{d,k}})^T, ~~~ 1\leq k \leq n_c,
\end{eqnarray}
and $\eta=10^{-2}$. The chosen values are the same as those considered in \cite{DeMaio-PAR}.
\end{itemize}
We let $\bR=\bM^{-1} \odot (\bp \bp^H)^*$ (see (\ref{eq:SNR}) and the following discussion) where $\bp$ is an unimodular vector with a structure similar to that of $\{ \bp_{v_{d,k}}\}$ in (\ref{eq:p_struct}). The UQP for the above cases is solved via MERIT using $20$ different random initializations for sizes $n \in \{ 8,16,32,64 \}$. Similar to the previous example, we also used SDR to solve the same UQPs. The results are shown in Table~\ref{table_cases}. The obtained solutions can be considered to be quite accurate in the sense of a sub-optimality guarantee $\gamma$ close to one. %We note that SDR can also solve the UQPs associated with the above cases quite accurately. 
 
\begin{table*}
\begin{center}
\footnotesize
\begin{tabular}{|c|c|p{2.0cm}|p{2.0cm}|p{2.0cm}|p{2.0cm}|p{2.0cm}|}
  \hline
  $n$ & Rank ($d$) & \#{problems for which $\gamma=1$} & {Average $\gamma$} & {Minimum $\gamma$} & {Average CPU time (sec)} & \#{problems solved by SDR} \\% & $\mathbb{E} \left[ \frac{v_{SDR}}{v_{MERIT}} \right]$\\

  \hline 
  \hline 
    & \textit{Case 1} & $20$& $1.0000$& $1.0000$& $2.82$ & $17$  \\ \cline{2-7}
   \textbf{8} & \textit{Case 2} & $20$& $1.0000$& $1.0000$& $0.60$& $20$ \\ \cline{2-7}
   & \textit{Case 3}& $20$& $1.0000$& $1.0000$& $0.27$ & $10$ \\ 
  \hline \hline
    & \textit{Case 1} & $20$& $1.0000$& $1.0000$& $42.83$ & $20$ \\ \cline{2-7}
 	\textbf{16} & \textit{Case 2} & $18$& $0.9812$& $0.8075$& $21.58$ & $20$ \\ \cline{2-7}
	 & \textit{Case 3}& $20$& $1.0000$& $1.0000$& $2.01$& $12$ \\ 
\hline \hline
     & \textit{Case 1} & $20$& $1.0000$& $1.0000$& $990.90$ & $20$ \\ \cline{2-7}
 	\textbf{32} & \textit{Case 2} &$19$& $0.9995$& $0.9913$& $525.34$ & $20$ \\ \cline{2-7}
	 & \textit{Case 3} & $20$& $1.0000$& $1.0000$& $7.52$ & $7$ \\
\hline \hline
    & \textit{Case 1} &$17$& $0.9901$& $0.9862$& $5574.98$ & $20$ \\ \cline{2-7}
 	\textbf{64} & \textit{Case 2} & $16$& $0.9540$& $0.8359$& $2053.26$ & $20$ \\ \cline{2-7}
	 & \textit{Case 3} & $20$& $1.0000$& $1.0000$& $22.78$ & $9$ \\	
  \hline
\end{tabular}
\end{center}
\caption{Comparison of the performance of MERIT (see Table~\ref{table:alpha_0_0}) and SDR \cite{DeMaio-PAR} when solving the UQP for the matrix structures described in Cases 1-3 using $20$ different initializations and for different sizes $n$.}
\label{table_cases}
\end{table*}

A different code design problem arises when 
synthesizing waveforms that have good resolution properties in range and Doppler \nocite{levanon}\nocite{seq_book}\nocite{unimodular_good}\nocite{soltanalian_comp_sec_design}[3]-[5],[24]-\cite{Phasecoded}. 
In the following, we consider the design of a thumbtack CAF (see the definitions in sub-section \ref{subsec:background}):
\begin{eqnarray}
d(\tau,f) = \left\{ \begin{array}{ll}
n & (\tau,f)=(0,0), \\
 0 & \mbox{otherwise.}
\end{array} \right.
\end{eqnarray}
Suppose $n=53$, let $T$ be the time duration of the total waveform, and let $t_p= T / n$ represent the time duration of each sub-pulse. Define the weighting function as
\begin{eqnarray}
w(\tau,f) = \left\{ \begin{array}{ll}
1 & (\tau,f)\in \Psi \backslash \Psi_{ml}, \\
 0 & \mbox{otherwise,}
\end{array} \right.
\end{eqnarray}
where $\Psi=[-10 t_p, 10 t_p] \times [-2/ T, 2/ T]$ is the region of interest and $\Psi_{ml}=([- t_p,  t_p] \backslash \{ 0\}) \times ([-1/ T, 1/ T]  \\ \backslash \{ 0\}) $ is the mainlobe area which is excluded due to the sharp changes near the origin of $d(\tau,f)$. Note that the time delay $\tau$ and the Doppler frequency $f$ are typically normalized by $T$ and $1/T$, respectively, and as a result the value of $t_p$ can be chosen freely without changing the performance of CAF design.  The synthesis of the desired CAF is accomplished via the cyclic minimization of (\ref{eq:CAF_crit}) with respect to $\bx$ and $\by$ (see sub-section \ref{subsec:background}). In particular, we use MERIT to obtain a unimodular $\bx$ in each iteration. A Bj\"orck code is used to initialize both vectors $\bx$ and $\by$.  The Bj\"orck code of length $n=p$ (where $p$ is a prime number for which $p \equiv 1~(\bmod~4)$) is given by
%\begin{eqnarray}
$\bb(k)=e^{j (\frac{k}{p}) \arccos\left(1/(1+\sqrt{p})\right)}$, $0 \leq k <p$,
%\end{eqnarray}
with $(\frac{k}{p})$ denoting the Legendre symbol. Fig. 2 depicts the normalized CAF modulus of the Bj\"orck code (i.e. the initial CAF) and the obtained CAF using the UQP formulation in (\ref{eq:CAF_UQP2}) and the proposed method. Despite the fact that designing CAF with a unimodular transmit vector $\bx$ is a rather constrained problem, MERIT is able to efficiently suppress the CAF sidelobes in the region of interest.

\begin{figure*}[tp] 
\centering
\subfigure[left: 3D plot, right: 2D plot]{\includegraphics[width=8.8cm]{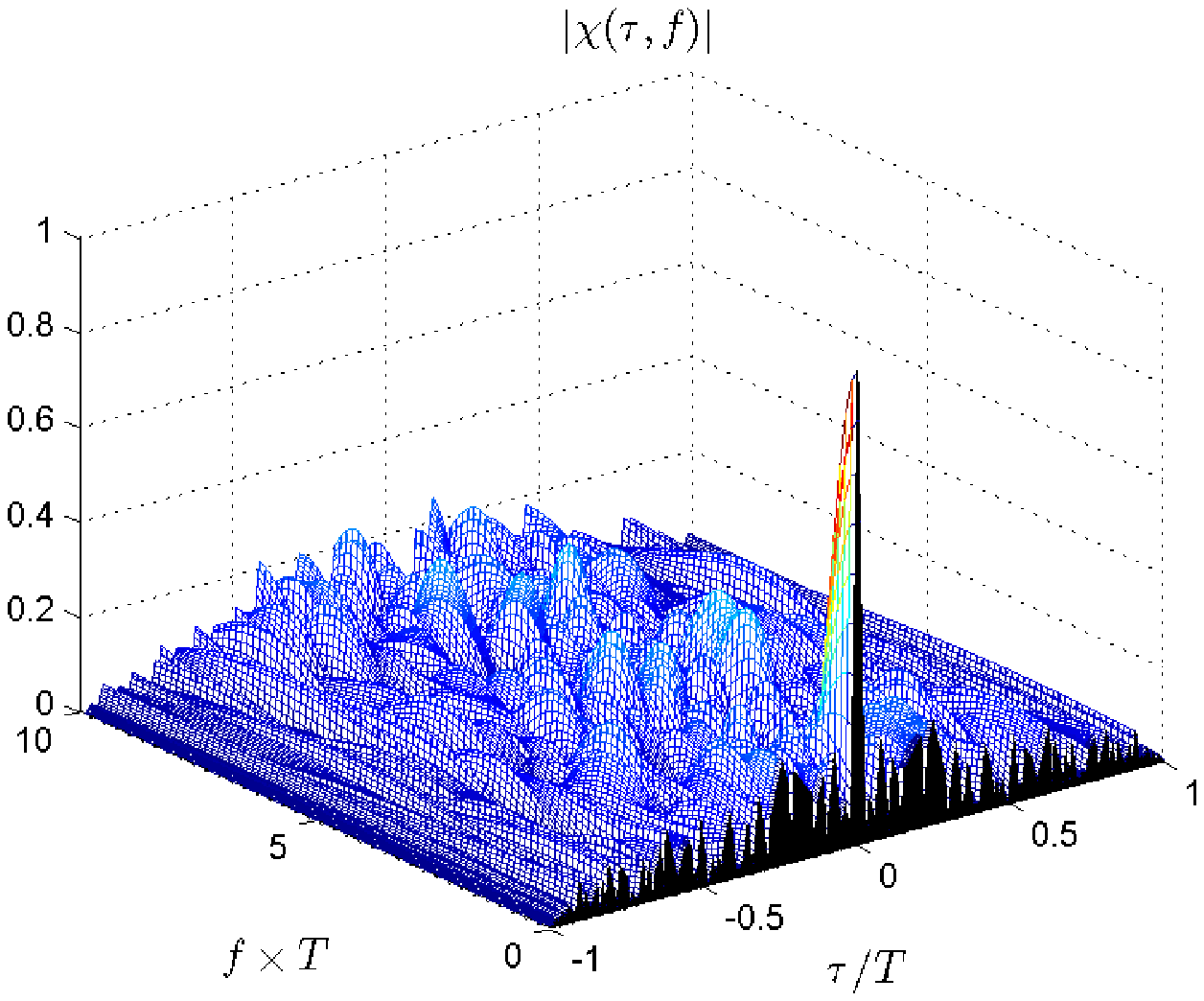}\label{fig:AF_bjork_3}\includegraphics[width=8.8cm]{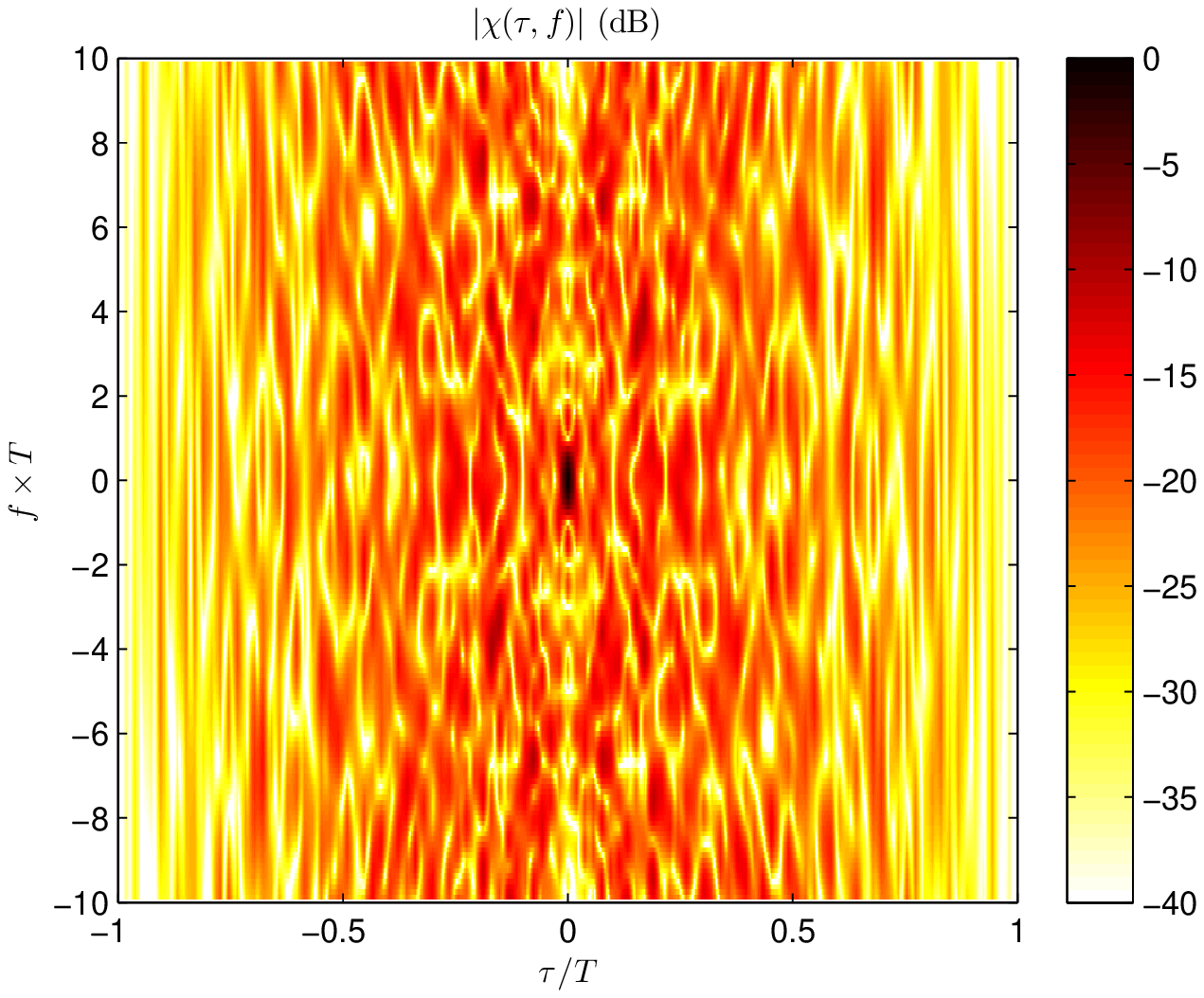}\label{fig:AF_bjork_2} 
}\\ \subfigure[left: 3D plot, right: 2D plot]{\includegraphics[width=8.8cm]{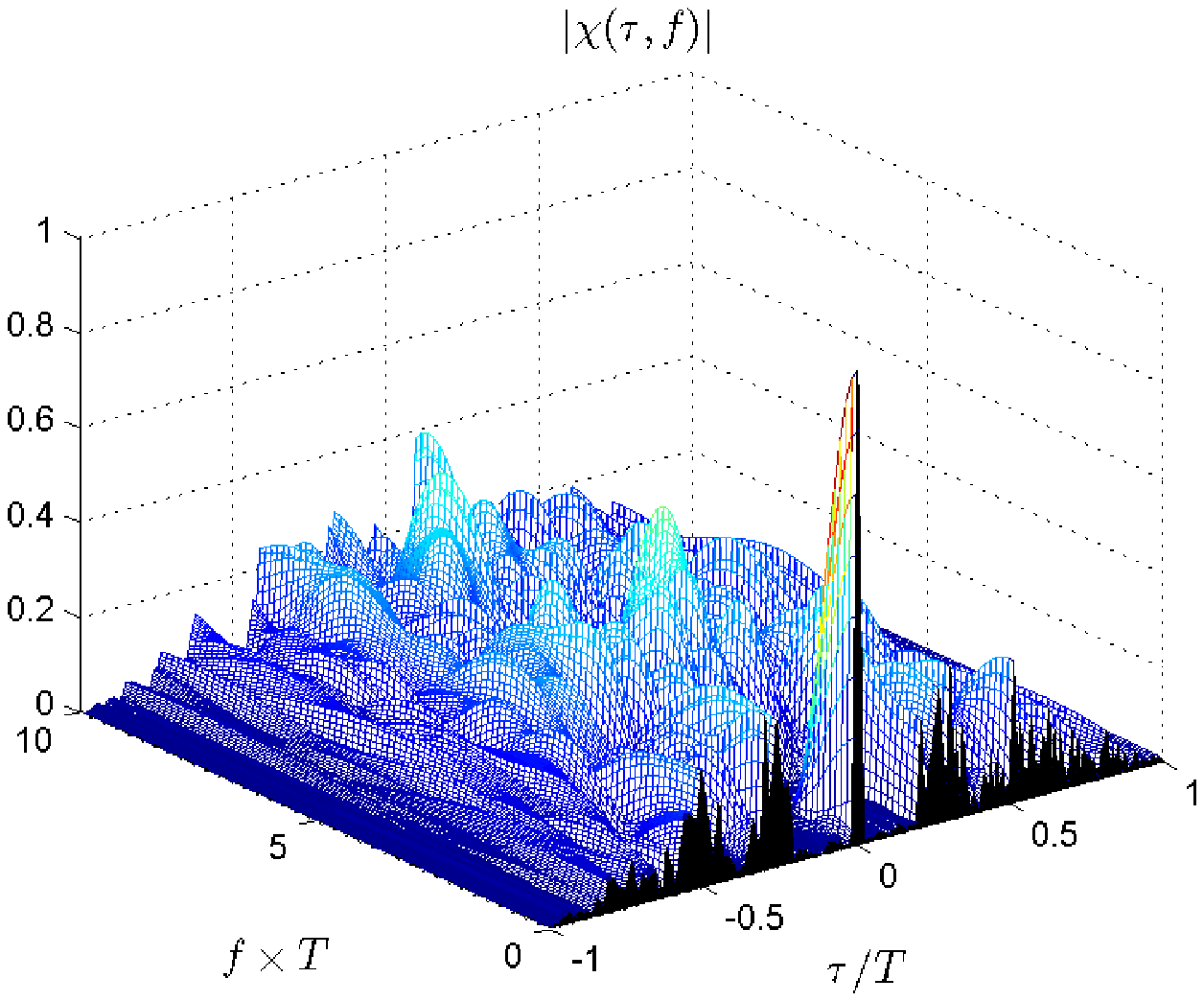}\label{fig:AF_s_3}\includegraphics[width=8.8cm]{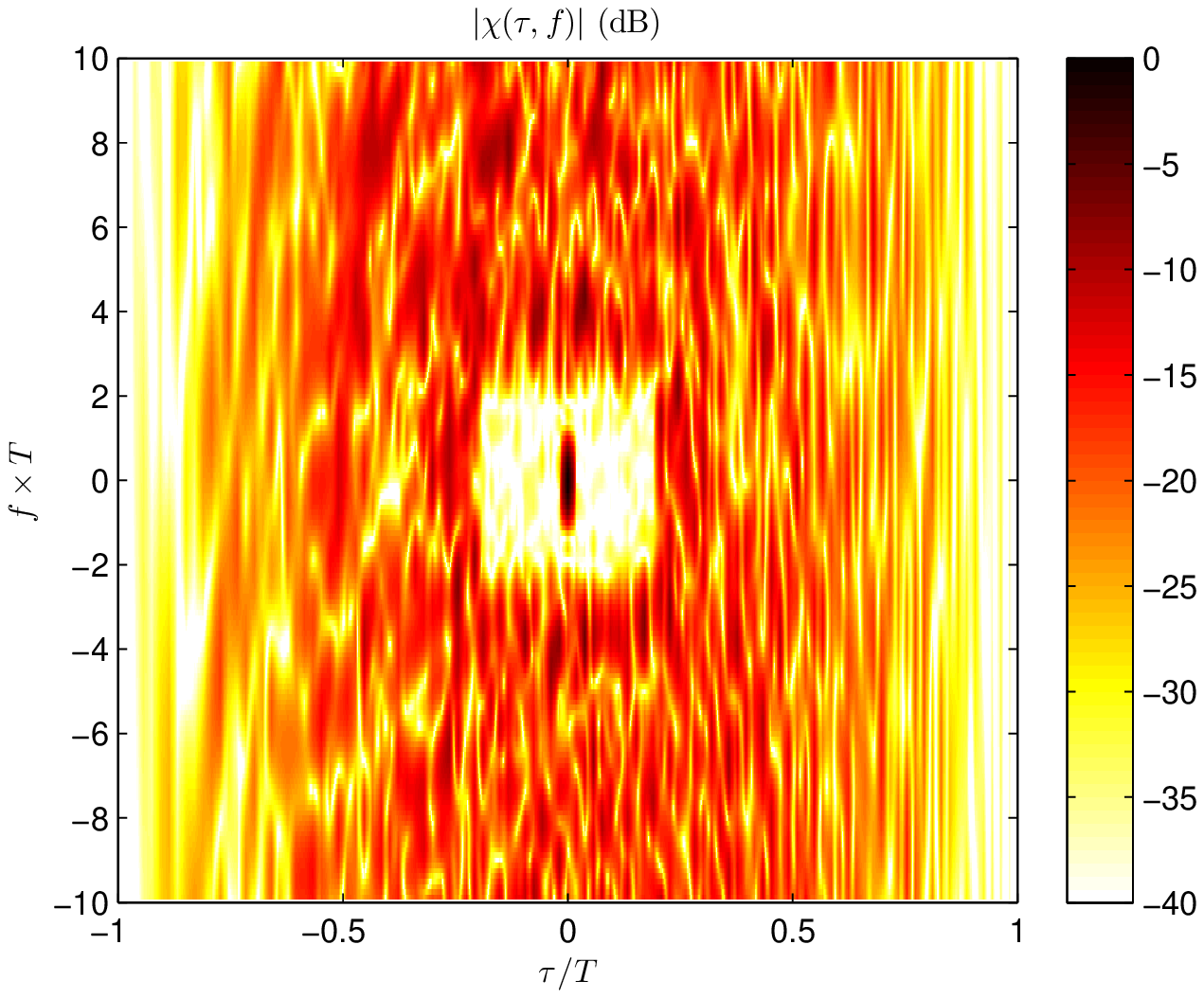}\label{fig:AF_s_2}
}
\caption{The normalized CAF modulus for (a) the Bj\"orck code of length $n=53$ (i.e. the initial CAF), and (b) the UQP formulation in (\ref{eq:CAF_UQP2}) and MERIT.}
\end{figure*}

\section{Concluding Remarks}\label{sec:conclusion}
A computational approach to the NP-hard problem of optimizing a quadratic form over the unimodular vector set (called UQP) has been introduced. The main results can be summarized as follows:
\begin{itemize}
\item Some applications of the UQP were reviewed. It was shown that the solution to UQP is not necessarily unique. Several examples were provided for which an  accurate global optimum of UQP can be obtained efficiently.
\item Using a relaxed version of UQP, a specialized local optimization scheme for UQP was devised and was shown to yield superior results compared to any general local optimization of UQP.
%It was shown that UQP is equivalent to a relaxed version of itself. This result was used to devise a local optimization scheme for UQP as well as characterize the local optima of UQP.
\item It was shown that the set of matrices ($\mathcal{K}(\bs)$) leading to the same solution ($\bs$) as the global optimum of UQP is a convex cone. An one-to-one mapping between any two such convex cones was introduced and an approximate characterization of $\mathcal{K}(\bs)$ was proposed.
\item Using the approximate characterization of $\mathcal{K}(\bs)$, an iterative approach (called MERIT) to the UQP was proposed. It was shown that MERIT provides case-dependent sub-optimality guarantees. The available numerical evidence shows that the sub-optimality guarantees obtained by MERIT are generally better than the currently known approximation guarantee (of $\pi /4$ for SDR).
\item Numerical examples were provided to examine the potential of MERIT for different UQPs. In particular, it was shown that the UQP solutions for certain matrices used in active sensing code design can be obtained efficiently via MERIT.
\end{itemize}

We should note that no theoretical efficiency assessment of the method was provided. It is clear that $\mathcal{C}(V_\bs) \cup \mathcal{C}_\bs \subset \mathcal{K}(\bs)$. A possible approach would be to determine how large is the part of $\mathcal{K}(\bs)$ that is ``covered" by $\mathcal{C}(V_\bs) \cup \mathcal{C}_\bs$. However, this problem is left for future work. Furthermore, a study of $m$-UQP using the ideas in this paper will be the subject of another paper.

%\section{Further Comments}\label{sec:comments}
%[An interesting phenomenon regarding UQP is that generally there exist many local optimas close to the global optima with objective values very close to the global optima as well. We can show this phenomenon by finding the best objective value and using our local optimizer to find many local optimas possessing mentioned property. Note that the importance of pinpointing this property is that it shows the inefficiency of currently known approximations ($\pi/4$ by [] and []).]

%[Which ratio of the volume of $\mathcal{K}(\bs)$ is covered by $\mathcal{C}(V_\bs) \cup \mathcal{C}_\bs$? I guess it is close to $1$. It is very good if we can calculate this ratio.]

\appendix
%\begin{center}
%Proofs of Theorems 1-3 and 6-8
%\end{center}

\subsection{Proof of Theorem 1} \label{subsec:th1}
In order to verify the first part of the theorem, consider any two matrices $\bR_1 ,\bR_2 \in \mathcal{K}(\widetilde{\bs})$. For any two non-negative scalars $\gamma_1, \gamma_2$ we have that
\begin{eqnarray}
\bs^H (\gamma_1 \bR_1 + \gamma_2 \bR_2) \bs = \gamma_1 \bs^H  \bR_1 \bs + \gamma_2 \bs^H  \bR_2 \bs.
\end{eqnarray}
Clearly, if some $\bs=\widetilde{\bs}$ is the global maximizer of both $\bs^H  \bR_1 \bs$ and $\bs^H  \bR_2 \bs$ then it is the global maximizer of $\bs^H (\gamma_1 \bR_1 + \gamma_2 \bR_2) \bs$ which implies $\gamma_1 \bR_1 + \gamma_2 \bR_2 \in \mathcal{K}(\widetilde{\bs}) $.

The second part of the theorem can be shown noting that
\begin{eqnarray}
\bs_2^H  (\bR \odot (\bs_0 \bs_0^H)) \bs_2 &=& (\bs_0^* \odot \bs_2)^H \bR  (\bs_0^* \odot \bs_2) \\ \nonumber
&=& \bs_1^H \bR \bs_1
\end{eqnarray}
for all  $\bs_1,\bs_2  \in \Omega^n$ and $\bs_0=\bs_1^* \odot \bs_2$. Therefore, if $\bR \in \mathcal{K}(\widetilde{\bs}_1)$ then $\bR \odot (\widetilde{\bs}_0 \widetilde{\bs}_0^H) \in   \mathcal{K}(\widetilde{\bs}_2)$ (for $\widetilde{\bs}_0=\widetilde{\bs}_1^* \odot \widetilde{\bs}_2$) and vice versa.

\subsection{Proof of Theorem 2} \label{subsec:th2}
It is well-known that 
%\begin{eqnarray}
$\bx^H \bR \bx \leq \sigma_1 \| \bx \|_2^2$  
%\end{eqnarray} 
 for all vectors $\bx \in \complexC^n$. Let 
\begin{eqnarray}
\balpha'=\left( \begin{array}{c}
\balpha \\
\bzero_{(n-m) \times 1}
\end{array} \right).
\end{eqnarray}
It follows from (\ref{eq:sualpha}) that $\widetilde{\bs} = \bU \balpha' $ and therefore
\begin{eqnarray}
\widetilde{\bs}^H \bR \widetilde{\bs} &=& \balpha'^H \bSig \balpha' = \sigma_1 \| \balpha' \|_2^2  \\ \nonumber
&=& \sigma_1 \| \widetilde{\bs} \|_2^2 = n  \sigma_1
\end{eqnarray}
which implies the global optimality of $\widetilde{\bs}$ for the considered UQP.

\subsection{Proof of Theorem 4} \label{subsec:th4}

It is worthwhile to observe that the convergence rate of $\{\bR^{(t)} \}$ is not dependent on the problem dimension ($n$), as each entry of $\{\bR^{(t)} \}$ is treated independently from the other entries (i.e. all the operations are element-wise). Therefore, without loss of generality we study the convergence of one entry (say $\{\bR^{(t)} (k,l) \}=\{ r_t e^{j \theta_t} \}$) in the following.

Note that in cases for which $|\theta_t - (\phi_k - \phi_l)| > \pi / 2$, the next element of the sequence $\{ r_t e^{j \theta_t} \}$ can be written as
\begin{eqnarray}
 r_{t+1} e^{j \theta_{t+1}} =  r_t e^{j \theta_t} + \rho e^{j(\phi_k - \phi_l)}
\end{eqnarray}
which implies that the proposed operation tends to make $\theta_t$ closer to $(\phi_k - \phi_l)$ in each iteration, and finally puts $\theta_t$ within the $ \pi / 2$ distance from $(\phi_k - \phi_l)$.

Let us suppose that $|\theta_0 - (\phi_k - \phi_l)| > \pi / 2$, and that the latter phase criterion remains satisfied for all $\theta_t$, $t<T$. We have that
\begin{eqnarray}
 r_{T} e^{j \theta_{T}} =  r_0 e^{j \theta_0} + T \rho e^{j(\phi_k - \phi_l)}
\end{eqnarray}
which yields
\begin{eqnarray}
 r_{T} \cos(\theta_T - (\phi_k - \phi_l))=  r_0 \cos(\theta_0 - (\phi_k - \phi_l)) + T \rho.
\end{eqnarray}
Therefore it takes only
%\begin{eqnarray}
$T=\left\lceil - r_0 \cos(\theta_0 - (\phi_k - \phi_l)) / \rho \right\rceil =1$
%\end{eqnarray}
iteration for $\theta_t$ to stand within the  $ \pi / 2$ distance from $(\phi_k - \phi_l)$.

Now, suppose that $|\theta_0 - (\phi_k - \phi_l)| \leq \pi / 2$. For every $t\geq 1$ we can write that 
\begin{eqnarray} \label{eq:rec}
 r_{t+1} e^{j \theta_{t+1}} &=&  r_t e^{j \theta_t} + \rho e^{j(\phi_k - \phi_l)}   \\ \nonumber &-&  r_t \cos(\theta_t - (\phi_k - \phi_l)) e^{j(\phi_k - \phi_l)} \\ \nonumber &=&e^{j(\phi_k - \phi_l)} \left( \rho + j  r_t \sin(\theta_t - (\phi_k - \phi_l)) \right).
 \end{eqnarray}
Let $\delta_{t+1}= r_{t+1} e^{j \theta_{t+1}} -  r_{t} e^{j \theta_{t}}$. The first equality in (\ref{eq:rec}) implies that
\begin{eqnarray} \label{eq:deltaarg1}
\delta_{t+1}=e^{j(\phi_k - \phi_l)} (\rho -  r_t \cos(\theta_t - (\phi_k - \phi_l)) )  .
\end{eqnarray}
On the other hand, the second equality in (\ref{eq:rec}) implies that
\begin{eqnarray} \label{eq:deltaarg2}
\delta_{t+1}= j e^{j(\phi_k - \phi_l)}  ( r_t \sin(\theta_t - (\phi_k - \phi_l))  ~~~~~~~~~~~~\\ \nonumber  - r_{t-1} \sin(\theta_{t-1} - (\phi_k - \phi_l)) )
\end{eqnarray}
for all $t\geq 1$. Note that in (\ref{eq:deltaarg1}) and (\ref{eq:deltaarg2}),  $\delta_{t+1}$ is a complex number having different phases. We conclude
\begin{eqnarray}
\delta_{t+1}=0, ~~ \forall ~t \geq 1
\end{eqnarray}
which shows that the sequence $\{ r_{t} e^{j \theta_{t}} \}$ is convergent in one iteration. In sum, every entry of the matrix $R$ will converge in at most two iterations (i.e. at most one to achieve a phase value within the  $ \pi / 2$ distance from $(\phi_k - \phi_l)$, and one iteration thereafter).

\subsection{Proof of Theorem 5} \label{subsec:th5}

We use the same notations as in the proof of Theorem 4. If $|\theta_0 - (\phi_k - \phi_l)| \leq \pi / 2$ then
\begin{eqnarray} \label{eq:lambda_diff1}
r_{2} e^{j \theta_{2}}&=& r_{1} e^{j \theta_{1}} \\ \nonumber &=&  r_0 e^{j \theta_0} + \rho e^{j(\phi_k - \phi_l)}   \\ \nonumber &-&  r_0 \cos(\theta_0 - (\phi_k - \phi_l)) e^{j(\phi_k - \phi_l)}.
\end{eqnarray}
On the other hand, if $|\theta_0 - (\phi_k - \phi_l)| > \pi / 2$ we have that
%\begin{eqnarray}
 $r_{1} e^{j \theta_{1}} =  r_0 e^{j \theta_0} +  \rho e^{j(\phi_k - \phi_l)}$.
%\end{eqnarray} 
As a result, $r_1 \cos(\theta_1 - (\phi_k - \phi_l)) = \rho + r_0 \cos(\theta_0 - (\phi_k - \phi_l)) $ which implies
\begin{eqnarray} \label{eq:lambda_diff2}
r_{2} e^{j \theta_{2}} &=&  r_1 e^{j \theta_1} + \rho e^{j(\phi_k - \phi_l)}   \\ \nonumber &-&  r_1 \cos(\theta_1 - (\phi_k - \phi_l)) e^{j(\phi_k - \phi_l)} 
\\ \nonumber &=&  r_0 e^{j \theta_0} + \rho e^{j(\phi_k - \phi_l)}   \\ \nonumber &-&  r_0 \cos(\theta_0 - (\phi_k - \phi_l)) e^{j(\phi_k - \phi_l)}.
\end{eqnarray}
Now, it is easy to verify that (\ref{eq:lambda_diff}) follows directly from (\ref{eq:lambda_diff1}) and (\ref{eq:lambda_diff2}).

\subsection{Proof of Theorem 6} \label{subsec:th6}

 If $\bs$ is a hyper point of UQP associated with $\bR^{(0)}=\bR$ then we have that 
%\begin{eqnarray} \label{eq:sglobal}
$\arg(\bs)=\arg(\bR \bs)$. 
%\end{eqnarray}
Let $\bR \bs= \bv \odot \bs$ where $\bv$ is a non-negative real-valued vector in $\realR^n$. It follows that
\begin{eqnarray}
\bv(k) e^{j \phi_k} = \sum_{l=1}^{n} |\bR (k,l)| e^{j \theta_{k,l}} e^{j \phi_l}
\end{eqnarray}
or equivalently
\begin{eqnarray}
\bv(k)  = \sum_{l=1}^{n} |\bR (k,l)| e^{j (\theta_{k,l}-( \phi_k - \phi_l))}
\end{eqnarray}
which implies that
\begin{eqnarray}
\left\{ \begin{array}{l}
\sum_{l=1}^{n} |\bR (k,l)| \cos( \theta_{k,l}-( \phi_k - \phi_l)) \geq 0 \\
\sum_{l=1}^{n} |\bR (k,l)| \sin( \theta_{k,l}-( \phi_k - \phi_l)) = 0 
\end{array} \right.
\end{eqnarray}
for all $1 \leq k \leq n$. Now, note that the recursive formula of the sequence $\{\bR^{(t)} \}$ can be rewritten as
\begin{eqnarray}
\bR^{(t+1)}=\bR^{(t)} - \mathbf{Diag} (\bs) ~(\bR_+^{(t)} - \rho \bone_{n \times n}) ~\mathbf{Diag}(\bs^*)
\end{eqnarray}
and as a result,
\begin{eqnarray} \label{eq:recursive}
\bR^{(t+1)} \bs = \bR^{(t)} \bs - \mathbf{Diag} (\bs) ~(\bR_+^{(t)} - \rho \bone_{n \times n}) ~\bone_{n \times 1}.
\end{eqnarray}
It follows from (\ref{eq:recursive}) that if $\bs$ is a hyper point of the UQP associated with $\bR^{(t)}$ (which implies the existence of non-negative real-valued vector $\bv^{(t)}$ such that $\bR^{(t)} \bs =  \bv^{(t)} \odot \bs$), then there exists  $\bv^{(t+1)} \in \realR^n$ for which $\bR^{(t+1)} \bs =  \bv^{(t+1)} \odot \bs$ and therefore,
\begin{eqnarray} \label{eq:55}
\bv^{(t+1)}(k) \, e^{j \phi_k} &=& \sum_{l=1}^{n} |\bR^{(t)} (k,l)| e^{j \theta_{k,l}} e^{j \phi_l} \\ \nonumber
 &-&  \left(  \left( \sum_{l=1}^{n} \bR^{(t)}_{+}(k,l) \right) - n \rho \right) e^{j \phi_k}.
\end{eqnarray}
Eq. (\ref{eq:55}) can be rewritten as
\begin{eqnarray} \label{eq:56}
\bv^{(t+1)} (k) &=& \sum_{l=1}^{n} |\bR^{(t)}(k,l)| e^{j (\theta_{k,l}-( \phi_k - \phi_l))} \\ \nonumber
 &-&  \left(   \sum_{l=1}^{n} \bR^{(t)}_{+}(k,l) \right) + n \rho  
\end{eqnarray}
As indicated earlier, $\bs$ being a hyper point for $\bR^{(0)}$ assures that the imaginary part of (\ref{eq:56}) is zero. To show that $\bs$ is a hyper point of the UQP associated with $\bR^{(t+1)}$, we only need to verify that $\bv^{(t+1)}(k) \geq 0$:
\begin{eqnarray}
\bv^{(t+1)}(k)&=& \sum_{l=1}^{n} |\bR^{(t)}(k,l)| \cos (\theta_{k,l}-( \phi_k - \phi_l)) \\ \nonumber
 &-&  \left(   \sum_{l=1}^{n} \bR^{(t)}_{+}(k,l) \right) + n \rho  \\ \nonumber
&=&   n \rho  \\ \nonumber  &+& \sum_{l:~(k,l) \notin \Theta}  |\bR^{(t)}(k,l)| \cos (\theta_{k,l}-( \phi_k - \phi_l))  
\end{eqnarray}
Now note that the positivity of $\bv^{(t+1)}(k)$ is concluded from (\ref{eq:lambda_ineq}). In particular, based on the discussions in the proof of Theorem 4, for  $t=1$, there is no $\theta_{k,l}$ such that $|\theta_{k,l} - (\phi_k - \phi_l)| \geq \pi / 2$ and therefore $\bv^{(2)}(k)= n \rho$ for all $1 \leq k \leq n$. As a result,
\begin{eqnarray}
\bR^{(2)} \bs =  n \rho  \bs
\end{eqnarray}
which implies that $\bs$ is an eigenvector of $\bR^{(2)}$ corresponding to the eigenvalue $n \rho$.

%\section{REFERENCES}

%\begin{eqnarray}
%\left\{ \begin{array}{l}
%\bR \bs_1 = \bV_2 \bs_2 \\
%\bR \bs_2 = \bV_1 \bs_1
%\end{array} \right.
%\end{eqnarray}
%where $\bV_1$ and $\bV_2$ are diagonal real-valued matrices which all their entries are non-negative (with at least one positive entry).

%Note that $\bs_1$ can be written as
%\begin{eqnarray} \label{eq:s1_decomp}
%\bs_1= a \bs_2 + b \bs'_2
%\end{eqnarray}
%where $a,b \in \complexC$ and $\bs'_2$ is a unit-vector perpendicular to $\bs_2$. Then
%\begin{eqnarray}
%\bR \bs_1= a \bR  \bs_2 + b \bR  \bs'_2
%\end{eqnarray}
%implying that
%\begin{eqnarray} \label{eq:s1_1}
%\bV_2 \bs_2= a \bV_1 \bs_1 + b \bR  \bs'_2.
%\end{eqnarray}
%Again, using (\ref{eq:s1_decomp}) we obtain that
%\begin{eqnarray} \label{eq:s1_2}
%a \bV_1 \bs_1= a^2  \bV_1 \bs_2 + a b  \bV_1 \bs'_2.
%\end{eqnarray}
%It follows from (\ref{eq:s1_1}) and (\ref{eq:s1_2}) that
%\begin{eqnarray}
%(\bV_2 -a^2  \bV_1)  \bs_2=  b( a  \bV_1 + \bR ) \bs'_2
%\end{eqnarray}
%Knowing that $\bs'_2$ is perpendicular to $\bs_2$, the latter equality implies that
%\begin{eqnarray}
%\left\{ \begin{array}{l}
%\bV_2 -a^2  \bV_1 = 0 \\
% b( a  \bV_1 + \bR )=0
%\end{array} \right.
%\end{eqnarray}
\section*{Acknowledgement}
We would like to thank Prof. Antonio De Maio for providing us with the MATLAB code for SDR.

\bibliographystyle{IEEEtran}
\bibliography{Des_uni_radarcodes_not_hard_1}

\begin{thebibliography}{10}
\providecommand{\url}[1]{#1}
\csname url@samestyle\endcsname
\providecommand{\newblock}{\relax}
\providecommand{\bibinfo}[2]{#2}
\providecommand{\BIBentrySTDinterwordspacing}{\spaceskip=0pt\relax}
\providecommand{\BIBentryALTinterwordstretchfactor}{4}
\providecommand{\BIBentryALTinterwordspacing}{\spaceskip=\fontdimen2\font plus
\BIBentryALTinterwordstretchfactor\fontdimen3\font minus
  \fontdimen4\font\relax}
\providecommand{\BIBforeignlanguage}[2]{{%
\expandafter\ifx\csname l@#1\endcsname\relax
\typeout{** WARNING: IEEEtran.bst: No hyphenation pattern has been}%
\typeout{** loaded for the language `#1'. Using the pattern for}%
\typeout{** the default language instead.}%
\else
\language=\csname l@#1\endcsname
\fi
#2}}
\providecommand{\BIBdecl}{\relax}
\BIBdecl

\bibitem{DeMaio-similarity}
A.~De~Maio, S.~De~Nicola, Y.~Huang, S.~Zhang, and A.~Farina, ``Code design to
  optimize radar detection performance under accuracy and similarity
  constraints,'' \emph{IEEE Transactions on Signal Processing}, vol.~56,
  no.~11, pp. 5618 --5629, Nov. 2008.

\bibitem{DeMaio-selection}
A.~De~Maio and A.~Farina, ``Code selection for radar performance
  optimization,'' in \emph{Waveform Diversity and Design Conference}, Pisa,
  Italy, June 2007, pp. 219--223.

\bibitem{seq_book}
H.~He, J.~Li, and P.~Stoica, \emph{Waveform Design for Active Sensing Systems:
  A Computational Approach}.\hskip 1em plus 0.5em minus 0.4em\relax Cambridge,
  UK: Cambridge University Press, 2012.

\bibitem{levanon}
N.~Levanon and E.~Mozeson, \emph{Radar Signals}.\hskip 1em plus 0.5em minus
  0.4em\relax New York: Wiley, 2004.

\bibitem{Hao_CAF}
H.~He, P.~Stoica, and J.~Li, ``On synthesizing cross ambiguity functions,'' in
  \emph{IEEE International Conference on Acoustics, Speech and Signal
  Processing (ICASSP)}, Prague, Czech Republic, May 2011, pp. 3536--3539.

\bibitem{Petre-beamformig}
J.~Li and P.~Stoica, \emph{Eds., Robust Adaptive Beamforming}.\hskip 1em plus
  0.5em minus 0.4em\relax NJ, USA.: John Wiley \& Sons, Inc., 2005.

\bibitem{steer-unique}
K.-C. Tan, G.-L. Oh, and M.~Er, ``A study of the uniqueness of steering vectors
  in array processing,'' \emph{Signal Processing}, vol.~34, no.~3, pp.
  245--256, 1993.

\bibitem{Khabbazibasmenj}
A.~Khabbazibasmenj, S.~Vorobyov, and A.~Hassanien, ``Robust adaptive
  beamforming via estimating steering vector based on semidefinite
  relaxation,'' in \emph{Conference on Signals, Systems and Computers
  (ASILOMAR)}, California, USA, Nov. 2010, pp. 1102--1106.

\bibitem{Jalden}
J.~Jalden, C.~Martin, and B.~Ottersten, ``Semidefinite programming for
  detection in linear systems - optimality conditions and space-time
  decoding,'' in \emph{IEEE International Conference on Acoustics, Speech, and
  Signal Processing (ICASSP)}, vol.~4, Hong Kong, April 2003, pp. 9--12.

\bibitem{zhang-complexquad}
S.~Zhang and Y.~Huang, ``Complex quadratic optimization and semidefinite
  programming,'' \emph{SIAM Journal on Optimization}, vol.~16, no.~3, pp.
  871--890, 2006.

\bibitem{low-rank}
A.~T. Kyrillidis and G.~N. Karystinos, ``Rank-deficient quadratic-form
  maximization over {M}-phase alphabet: Polynomial-complexity solvability and
  algorithmic developments,'' in \emph{IEEE International Conference on
  Acoustics, Speech and Signal Processing (ICASSP)}, May 2011, pp. 3856--3859.

\bibitem{sergio-multiusercomplexity}
S.~Verd\'u, ``Computational complexity of optimum multiuser detection,''
  \emph{Algorithmica}, vol.~4, pp. 303--312, 1989.

\bibitem{MaVo-blindML}
W.-K. Ma, B.-N. Vo, T.~Davidson, and P.-C. Ching, ``Blind {ML} detection of
  orthogonal space-time block codes: efficient high-performance
  implementations,'' \emph{IEEE Transactions on Signal Processing}, vol.~54,
  no.~2, pp. 738--751, feb. 2006.

\bibitem{Cui_OFDM_spheredecoing}
T.~Cui and C.~Tellambura, ``Joint channel estimation and data detection for
  {OFDM} systems via sphere decoding,'' in \emph{IEEE Global Telecommunications
  Conference (GLOBECOM)}, vol.~6, Texas, USA, Dec. 2004, pp. 3656--3660.

\bibitem{convex_boyd}
S.~Boyd and L.~Vandenberghe, \emph{{Convex Optimization}}.\hskip 1em plus 0.5em
  minus 0.4em\relax Cambridge, UK: Cambridge University Press, 2004.

\bibitem{Goemans}
M.~X. Goemans and D.~P. Williamson, ``Improved approximation algorithms for
  maximum cut and satisfiability problems using semidefinite programming,''
  \emph{ACM Journal}, vol.~42, no.~6, pp. 1115--1145, Nov. 1995.

\bibitem{DeMaio-PAR}
A.~De~Maio, Y.~Huang, M.~Piezzo, S.~Zhang, and A.~Farina, ``Design of optimized
  radar codes with a peak to average power ratio constraint,'' \emph{IEEE
  Transactions on Signal Processing}, vol.~59, no.~6, pp. 2683--2697, June
  2011.

\bibitem{SDR_mag}
Z.~Q. Luo, W.~K. Ma, A.-C. So, Y.~Ye, and S.~Zhang, ``Semidefinite relaxation
  of quadratic optimization problems,'' \emph{IEEE Signal Processing Magazine},
  vol.~27, no.~3, pp. 20 --34, May 2010.

\bibitem{so-approxSDR}
A.~So, J.~Zhang, and Y.~Ye, ``On approximating complex quadratic optimization
  problems via semidefinite programming relaxations,'' \emph{Mathematical
  Programming}, vol. 110, pp. 93--110, 2007.

\bibitem{nearest-doubly-stochastic}
W.~Glunt, T.~L. Hayden, and R.~Reams, ``The nearest 'doubly stochastic' matrix
  to a real matrix with the same first moment,'' \emph{Numerical Linear Algebra
  with Applications}, vol.~5, no.~6, pp. 475--482, 1998.

\bibitem{horn1990matrix}
R.~Horn and C.~Johnson, \emph{Matrix Analysis}.\hskip 1em plus 0.5em minus
  0.4em\relax Cambridge, UK: Cambridge University Press, 1990.

\bibitem{rankdef_binary}
G.~Karystinos and A.~Liavas, ``Efficient computation of the binary vector that
  maximizes a rank-deficient quadratic form,'' \emph{IEEE Transactions on
  Information Theory}, vol.~56, no.~7, pp. 3581--3593, July 2010.

\bibitem{Demaio-maxmin}
A.~De~Maio, Y.~Huang, and M.~Piezzo, ``A {Doppler} robust max-min approach to
  radar code design,'' \emph{IEEE Transactions on Signal Processing}, vol.~58,
  no.~9, pp. 4943--4947, Sept. 2010.

\bibitem{unimodular_good}
P.~Stoica, H.~He, and J.~Li, ``New algorithms for designing unimodular
  sequences with good correlation properties,'' \emph{IEEE Transactions on
  Signal Processing}, vol.~57, no.~4, pp. 1415--1425, April 2009.

\bibitem{soltanalian_comp_sec_design}
M.~Soltanalian and P.~Stoica, ``Computational design of sequences with good
  correlation properties,'' \emph{IEEE Transactions on Signal Processing},
  vol.~60, no.~5, pp. 2180--2193, May 2012.

\bibitem{Phasecoded}
J.~Benedetto, I.~Konstantinidis, and M.~Rangaswamy, ``Phase-coded waveforms and
  their design,'' \emph{IEEE Signal Processing Magazine}, vol.~26, no.~1, pp.
  22--31, Jan. 2009.

\end{thebibliography}

\end{document}